\documentclass[a4paper,11pt,reqno]{article}

\usepackage{graphicx,epstopdf}
\usepackage{amsmath, amsfonts, amssymb}
\usepackage[english]{babel}
\usepackage[breaklinks=true]{hyperref}
\usepackage{lpic}
\usepackage{nicefrac}
\usepackage[sort&compress, merge,numbers]{natbib}

\usepackage{booktabs}

\advance\textwidth by 1.4in\advance\oddsidemargin by - 0.7in
\advance\textheight by 1.1in\headheight 0pt\topskip 0pt\headsep 0pt\topmargin 0pt

\hypersetup{
			colorlinks=true,
			urlcolor=blue,
			citecolor=magenta,
			linkcolor=blue,
     }
\numberwithin{equation}{section}

\newcommand{\quoting}[1]{``#1''}
\newcommand{\td}{\text{d}}

\let\oldsqrt\sqrt
\def\sqrt{\mathpalette\DHLhksqrt}
\def\DHLhksqrt#1#2{%
\setbox0=\hbox{$#1\oldsqrt{#2\,}$}\dimen0=\ht0
\advance\dimen0-0.2\ht0
\setbox2=\hbox{\vrule height\ht0 depth -\dimen0}%
{\box0\lower0.4pt\box2}}

\begin{document}

\hypersetup{pageanchor=false}

\begin{center}

\vspace{.7cm}
{\LARGE\bf Probing the Hydrodynamic Limit of (Super)gravity} \\

\vspace{1.5cm}

{{\textbf{Adriana Di Dato}$^1$, \textbf{Jakob Gath}$^{2}$, \textbf{Andreas Vigand Pedersen}$^{3,4}$}\\
\vspace{1cm}
{$^1$ \small\slshape Departament de F\'{i}sica Fonamental, Institut de Ci\`{e}ncies del Cosmos,\\
 Universitat de Barcelona, C/ Mart\'{i} i Franqu\`{e}s 1, 08028, Barcelona, Spain.}\\
\vspace{0.4cm}

{$^2$ \small\slshape Centre de Physique Th\'{e}orique, Ecole Polytechnique, \\
CNRS UMR 7644, 91128 Palaiseau Cedex, France.}\\
\vspace{0.4cm}

{$^3$ \small\slshape Center for Theoretical Physics and Department of Physics, \\
 University of California, Berkeley, CA 94720, U.S.A.}\\
 \vspace{0.4cm}

{$^4$ \small\slshape Niels Bohr Institute, University of Copenhagen,\\
 Blegdamsvej 17, DK-2100 Copenhagen \O, Denmark.}}\\ 
\vspace{1.0cm}

{\small  \texttt{adidato@ffn.ub.es, gath@cpht.polytechnique.fr, vigand@berkeley.edu}}
\vspace{1.5cm}


\thispagestyle{empty}

{\bf Abstract} \end{center} \vspace{0mm} {
We study the long-wavelength effective description of two general classes of charged dilatonic (asymptotically flat) black $p$-branes including D/NS/M-branes in ten and eleven dimensional supergravity. 
In particular, we consider gravitational brane solutions in a hydrodynamic derivative expansion (to first order) for arbitrary dilaton coupling and for general brane and co-dimension and determine their effective electro-fluid-dynamic descriptions by exacting the characterizing transport coefficients.
%
We also investigate the stability properties of the corresponding hydrodynamic systems by analyzing their response to small long-wavelength perturbations.
For branes carrying unsmeared charge, we find that in a certain regime of parameter space there exists a branch of stable charged configurations. This is in accordance with the expectation that D/NS/M-branes have stable configurations, except for the D5, D6, and NS5. 
In contrast, we find that Maxwell charged brane configurations are Gregory-Laflamme unstable independently of the charge and, in particular, verify that smeared configurations of D0-branes are unstable. 
%
Finally, we provide a modification to the mapping presented in
\href{http://arxiv.org/abs/1211.2815}{arxiv:1211.2815}
and utilize it to provide a non-trivial cross-check on a certain subset of our transport coefficients with the results of 
\href{http://arxiv.org/abs/1110.2320}{arXiv:1110.2320}.
}

\newpage
\clearpage
\hypersetup{pageanchor=true}
\pagenumbering{arabic}

\noindent\rule{\textwidth}{1.2pt}
\vspace{-1cm}
\begingroup
\hypersetup{linkcolor=black}
\tableofcontents
\endgroup
\noindent\rule{\textwidth}{1.2pt}

\section*{Introduction}
\addcontentsline{toc}{section}{Introduction}


Charged gravitational string and brane solutions play an important r\^{o}le in string theory. A large class of these objects has for a long time been known to be the supergravity incarnations of the fundamental string and D$p$-branes \cite{Polchinski:1995mt}. Indeed, these spatially extended geometries source a set of gauge potentials and are solutions of (an appropriate) supergravity that are exactly interpreted as the classical coherent state of a (large) stack of fundamental strings/branes each charged under the corresponding potentials. This simple, yet profoundly deep, observation lies at the very heart of the AdS/CFT correspondence  \cite{Maldacena:1997re}. In general, the weakly coupled low-energy effective dynamics of the F-string and the D$p$-branes is well-understood and conventionally captured by the open string sector, i.e., the Nambu-Goto and Dirac-Born-Infeld (DBI) action, respectively. However, at strong coupling, the dynamics is more appropriately captured by the closed string sector \emph{viz.} supergravity. It is thus natural to ask how the various aspects of the low-energy DBI descriptions of the effective dynamics translate to the supergravity picture. The framework to answer these questions is, in part, provided by relativistic hydrodynamics. Indeed, the existence of such a connection is not far fetched since hydrodynamics is the natural local generalization of the global thermodynamics of black branes.

The effective description of gravity as a hydrodynamic theory is of course by no means new. The reader might be familiar with the membrane paradigm \citep{Damour:1978cg, *Price:1986yy} where the effective black \emph{hole} dynamics is formulated in terms of a non-relativistic fluid that lives on the (stretched) horizon of the black hole. Perhaps more familiar is the celebrated fluid/gravity correspondence \citep{Bhattacharyya:2008jc} which emerges as the natural hydrodynamic limit of the AdS/CFT correspondence. In its original formulation, the long-wavelength effective description of $\mathcal N=4$ SYM is identified with a relativistic hydrodynamic theory. Here the transport coefficients (for a review on relativistic fluid mechanics, see \citep{landau_fluid_1987}) are directly computed from gravity using linear response theory in \citep{Policastro:2001yc} and from a direct perturbative gravitational computation in \citep{Bhattacharyya:2008jc}. Moreover, various generalizations including charged cases, have been carried out in these schemes, see e.g. \citep{Maeda:2006by, *Son:2006em, *Banerjee:2008th, *Erdmenger:2008rm, *Bhattacharyya:2008ji}. The latter approach is quite similar in spirit to the one considered in the present work and has previously been applied to the Schwarzschild black $p$-brane in \citep{Camps:2010br} and for the Reissner-Nordstr\"om black brane in \citep{Gath:2013qya}. However, we emphasize that in this work we do not consider asymptotically (co-dimension 1) AdS branes but rather asymptotically flat configurations of general co-dimension.

In more detail, in this paper we shall be interested in the effective behavior of a quite general class of black brane solutions captured by the action \eqref{action} which will be introduced in Sec.~\ref{pequalqhydro}. Although this action is rather general, in this work, we will restrict ourselves to the cases where the black $p$-branes source a $(p+1)$-form \emph{or} a $1$-form gauge potential. In the following we will refer to these two types of brane charge as fundamental charge and Maxwell charge, respectively. In particular, the treatment of these two cases will allow us to capture the effective hydrodynamic descriptions of the NS and D$p$-branes of type II supergravity along with the M$2$ and M$5$-branes of eleven dimensional supergravity. Moreover, our computation also captures the effective theory of $(p+1)$-dimensional smeared brane configurations of D$0$ of type IIA supergravity. However, instead of fixing the value of the dilaton coupling and spacetime dimension to the ones relevant for the specific supergravities, we will keep these parameters free. This allows us to extract the dependence of these parameters in the hydrodynamic transport coefficients, which turns out to be quite useful for examining the general features of the hydrodynamics. We note that the part of the computation pertaining to Maxwell charge also provides the non-trivial dilatonic generalization of the results for the Reissner-Nordstr\"om black brane worked out in \cite{Gath:2013qya}.

\vspace{0.5cm}

Considerable research has gone into understanding the stability properties of black branes carrying various types of smeared charges (the literature is extensive, for a review see e.g. \citep{Harmark:2007md}). Understanding these properties is interesting from a pure general relativistic point of view \emph{viz.} the Gregory-Laflamme (GL) instability \citep{Gregory:1993vy, *Gregory:1994bj}, black hole phase transitions \citep{Kol:2002xz} etc., but also plays an important r\^{o}le for understanding aspects of the (un)stable vacua of string and M-theory. In general, having access to an effective fluid dynamic description naturally allows one to address questions regarding the stability properties of the system in question. For example, to leading order, i.e., at the perfect fluid level (equivalently; at the thermodynamic level), an imaginary speed of sound signifies a fundamental instability in the system. In the context of brane physics, this was noted in \citep{Emparan:2009at} where an instability in the sound mode of the effective fluid of the Schwarzschild black brane exactly was identified with the GL instability of the brane. A GL instability is naturally characterized by a dispersion relation which describes the \quoting{dispersion} of the instability on the worldvolume. Although the exact form of this relation is only accessible numerically many of its features can be understood from the hydrodynamic description. In particular, turning the picture around, the lack of unstable hydrodynamic modes in the effective theory (to a given order in the derivative expansion) is equivalent to the lack of a GL instability, at least to that given order.



One of the main results of this work is that (to first order in the fluid derivative expansion) the stability properties of the fundamentally charged dilatonic black brane remains solely determined by the speed of sound. In this way, the system will exhibit a stable branch of configurations for sufficiently small values of the dilaton coupling. This includes the D$p$-branes, $p<5$, of type II supergravity along with the M2 and M5 branes of M-theory. However, it does \emph{not} include the D5, D6 and NS5 brane. We note that a similar conclusion was reached in \cite{Emparan:2011hg}, but we refine the analysis to first order and use the correct value for the bulk viscosity. This result is furthermore in accordance with expectations from various numerical works \cite{Reall:2001ag, Gregory:1993vy, Gregory:1994bj}. On the other hand, we find that the dilatonic Maxwell charged black branes cannot be made stable for any value of the dilaton coupling or the charge. In particular, this includes the smeared D0 configurations of type IIA supergravity. Again, this is in accordance with general expectations \citep{Bostock:2004mg, *Aharony:2004ig}. We note that, as in Ref. \cite{Gath:2013qya}, this is a genuine first-order derivative effect, meaning that the instability is not visible to leading order. 

Another interesting aspect of our computation relates to various proposed hydrodynamic bounds in gravity \citep{Policastro:2001yc, Kovtun:2003wp, *Kovtun:2004de, *Buchel:2003tz, Buchel:2007mf}. As mentioned above, we keep the dilaton coupling free in our computations. In addition to elucidating the (in)stability properties of the various branes this also provides us with an extra turnable parameter to examine the possible violation of the hydrodynamic bounds. As expected, we find that the shear viscosity-to-entropy ratio bound $\eta/s\geq 1/4\pi$ is saturated for our entire class of gravitational solutions. In contrast we find that the (holographic) bulk viscosity \quoting{bound} is violated for the entire class of fundamentally charged branes thus providing a simple example of its violation.

\vspace{0.5cm}
  
Although the setup employed in this work is conceptually quite different from the hydrodynamic limit of AdS/CFT (the fluid/gravity correspondence), some of the results of the two approaches can nevertheless be related. This is perhaps not too surprising since the latter should in principle be obtainable from the former in the near-horizon limit. The precise connection between the effective fluid dynamic theory of the asymptotically flat (non-dilatonic) D3-brane and the fluid/gravity correspondence on $\text{AdS}_5$ was explored in Ref. \cite{Emparan:2013ila}. Here the effective gravitational dynamics of the D3-brane subject to Dirichlet boundary conditions at an appropriate cutoff surface was considered along the lines of ideas introduced in the paper(s) \citep{Bredberg:2010ky, *Bredberg:2011jq}. It was directly shown that the fluid/gravity correspondence  and the hydrodynamic effective theory are obtained as the two (most interesting) extremes where the cutoff surface is taken to be located in the near-horizon throat region and at spatial infinity, respectively. Perhaps more surprisingly, the results of the fluid/gravity correspondence can also be related to the effective hydrodynamics of the (neutral) Schwarzschild black $p$-brane. This was shown in the  paper(s) \citep{Caldarelli:2012hy, *Caldarelli:2013aaa} where the authors managed to derive a mapping from a certain class of asymptotically AdS solutions to asymptotically flat solutions and vice versa. This class of solutions (on each side of the mapping, respectively) exactly includes the Ricci-flat and the AdS Schwarzschild black branes, respectively. More technically, the mapping is established by noting that the hydrodynamic sector on either side (of the mapping) is completely included in a reduced lower dimensional theory. Solutions to these two reduced theories can then be related by a simple analytical continuation in the dimension of the particular space on which the reduction is performed (a sphere and a torus, respectively). In this way, solutions on one side can be reduced, analytically continued and consistently uplifted to the other side and vice versa. This, conceptually quite simple, procedure neatly allowed \citep{Caldarelli:2012hy} to derive the second-order fluid dynamic transport coefficients of the asymptotically flat Schwarzschild black brane from the results of the (co-dimension one) AdS black brane in general dimensions \citep{Bhattacharyya:2008mz, *Haack:2008cp}. In this paper we will present a modified version of the mapping which allows us to include Maxwell charge. 

The paper is organized as follows. In the Sec.~\ref{pequalqhydro} we introduce the leading order (seed) geometries around which we will consider hydrodynamic perturbations. Before presenting the first-order results of the fundamentally charged black brane and analyzing their stability, we briefly review the associated thermodynamics and explain how the perturbative procedure is carried out. In Sec.~\ref{pequal0hydro} we turn our attention to Maxwell charged black branes and work out the first-order transport coefficients and discuss the associated hydrodynamic stability properties. We conclude the paper by performing a non-trivial cross-check of our results with known results from AdS by introducing a modified version of the \quoting{AdS/Ricci-flat correspondence}.


\section{Hydrodynamics of black \texorpdfstring{$p$}{p}-branes }\label{pequalqhydro}

In this work we shall be interested in $D$ dimensional $p$-brane solutions (i.e., solutions with horizon topology $\mathbf R^p\times S^{D-p-2}$) of the following action\footnote{We work in units where $G_{\text{Newton}}=\frac{1}{16\pi}$.} \citep{Argurio:1997gt, *Argurio:1997nh}
\begin{equation}\label{action}
I=\int_D \! \left(R\ast 1 - 2 \thinspace \td \phi \wedge \ast \thinspace \td \phi -\frac{1}{2}\sum_{q\in\mathcal I} \mathcal F_{(q+2)}\wedge \ast \mathcal F_{(q+2)}\right) \ .
\end{equation}
Here $\mathcal F_{(q+2)}=e^{a_q\phi} \thinspace \td C_{(q+1)}$, $q\in \mathcal I$, are the dilaton weighted field strengths associated with the gauge potentials $C_{(q+1)}$ and $\mathcal I$ denotes the collective set of gauge potentials in the theory. Notice that some of the forms can be of the same rank nonetheless they have different couplings $a_q$ to the dilaton which distinguishes them. The action \eqref{action} is quite general, however, for specific choices of $\mathcal I$ and dilaton couplings, it captures the (bosonic part of the) supergravity descriptions of type IIA/B (in the Einstein frame) and M-theory relevant for description of the D/NS/M-branes and their intersections. Notice that the aforementioned supergravity actions also contain a topological term needed to preserve supersymmetry of the full theory, however, this term does not play a r\^{o}le for obtaining the flat $p$-brane solutions \cite{Peet:2000hn}. The action \eqref{action} corresponds to IIA (IIB) supergravity for $D=10$, $\mathcal I_{\text{RR}}=\{0,2\}$ ($\mathcal I_{\text{RR}}=\{ 1,3 \}$) and $\mathcal I_{\text{NS}}=\{1\}$ \citep{Huq:1983im, *Giani:1984wc, *Campbell1984112} and eleven-dimensional supergravity for $D=11$ and $\mathcal I_{\text{M}}=\{ 2\}$ \cite{Cremmer:1978km}. In general, the D/NS/M-branes couple both electrically and
magnetically to the above potentials. We unify the description in the
standard way by writing the field strengths in the electric ansatz, where now the index
$q$ in the action \eqref{action} runs over the (allowed) spatial dimensions of the branes of the theory. Given a dilaton
coupling $a$, it will be convenient to define a parameter $N$ to further unify the
description through the relation
\begin{equation}\label{asquaredef}
a^2=\frac{4}{N}-\frac{2(q+1)(D-q-3)}{D-2} \ .
\end{equation}
The real positive parameter $N$ (usually an integer for string/M-theory corresponding to the number
of different types of branes in an intersection \citep{Tseytlin:1996cc}) in general is preserved under dimensional reduction \citep{Caldarelli:2010xz}.\footnote{Since we require the dilaton to be physical (i.e., we require $a^2\geq 0$), the parameter $N$ is bounded from above $N \leq \frac{2(D-2)}{(D-q-3)(q+1)}$.}
%
%
Also note that the parameter $N$ mods out the $\mathbf Z_2$ reflection symmetry of the solution space $a\to-a$. For all the fundamental D/NS/M-branes of type II string theory and M-theory $N = 1$, and $a_{\text{D}p} = (3-p)/2$
for the D$p$-branes, while $a^2 = 1$ for the F1 ($a_{\text{F}1} = -1$) and NS5 brane ($a_{\text{NS}5} = 1$). Also
notice that the dilaton coupling \eqref{asquaredef} vanishes for $D = 11$, $q = 2, 5$, consistent with the
fact that M-theory contains no dilaton.

In the following we will consider singly fundamentally charged $p$-branes, i.e., $q=p$. We therefore (consistently) truncate the action \eqref{action} to $\mathcal I=\{ p \}$. Moreover, we will keep the dilaton coupling $a\equiv a_p$ free. One easily works out the equations of motion, 
\begin{align}\label{EOMsip}
\begin{split}
\square \phi &= \frac{a}{4(p+2)!} \thinspace \mathcal F^2 \ , \quad
\td \left( e^{a\phi} \ast \mathcal F  \right) =0 \ , \\
\thinspace G^{\nu}_{\ \mu} &= \frac{1}{2(p+1)!} \left(\mathcal F\cdot \mathcal F\right)^{\nu}_{\ \mu} + \left( 2\left(\partial\phi\right)^2 - \frac{ \thinspace \mathcal F^2 }{4(p+2)!} \thinspace \right) \delta^\nu_{\ \mu} \ ,
\end{split}
\end{align}
where $G_{\mu\nu}$ denotes the Einstein tensor and $\mu,\nu$ label the spacetime directions. As explained in the introduction, we are interested in $p$-brane solutions to the theory \eqref{action} characterized by horizon topology $\mathbf R^p \times S^{n+1}$, where we break the $\partial_i$ symmetries on $\mathbf R^p$ (and thus implicitly breaking  $\partial_t$ as well) perturbatively while maintaining the $\text{SO}(n+2)$ symmetry on the transverse sphere (note that the total spacetime dimension $D$ is related to the positive integer parameter $n$ according to $n=D-p-3$). These perturbations exactly capture the hydrodynamic sector of the black brane as we review below.\footnote{Breaking the $\text{SO}(n+2)$ would roughly correspond to elastic perturbations, see \cite{Armas:2011uf} and related works.} The family of leading order unperturbed (seed) $p$-branes solutions to the EOMs \eqref{EOMsip} was worked out in Ref. \cite{Caldarelli:2010xz} and takes the form
\begin{equation}\label{solutionSCH}
\begin{split}
\td s^2&=h^{-\frac{Nn}{p+n+1}}\Bigg(-f u_a u_b \thinspace \td x^a \td x^b+\Delta_{ab} \thinspace\td x^a \td x^b + h^N\left(f^{-1}\td r^2+r^2 \thinspace \td \Omega_{(n+1)}^2\right) \Bigg) \ , \\ 
&\ \ \quad \phi=\frac{aN}{4}\log h \ , \quad \quad A_{(p+1)}= -\sqrt{N\left(\frac{\gamma_0+1}{\gamma_0}\right)} \thinspace \left(h^{-1}-1\right)\star  1 \ .
\end{split}
\end{equation} 
Here we have applied a general boost $u^a$ ($u_a u^a=-1$) to the solution in the $p+1$ worldvolume directions labeled by $x^a=(t,x^i)$. The tensor $\Delta_{ab}=\eta_{ab}+u_a u_b$ is the projector onto the directions parallel to the brane but orthogonal to $u^a$, while $\star 1=\td x^0\wedge \dotsc \wedge \td x^p$ denotes the induced volume form on the brane geometry. The two functions $f$ and $h$ are given by,
\begin{equation}
f(r)\equiv 1-\left(\frac{r_0}{r}\right)^n \ , \quad 
h(r)\equiv  1+\gamma_0 \left(\frac{r_0}{r}\right)^n \ .
\end{equation}
Here $r_0$ parameterizes the horizon radius and $\gamma_0$ parameterizes the charge of the solution.

According to fluid/gravity lore there is a one-to-one correspondence between the solutions to the EOMs \eqref{EOMsip} around the solution \eqref{solutionSCH}, and the relativistic Navier-Stokes equations, 
\begin{equation} \label{eqn:conservation}
\text{div} \thinspace T=0 \ , \quad \td \star j=0 \ .
\end{equation}
Here $T=T^{ab}$ is the effective stress tensor and $j$ is the effective current which are matched order by order in a relativistic fluid derivative expansion. The effective stress tensor and current encompass the asymptotic data of the perturbed solution and the correspondence allows one to reconstruct the full gravitational solution (to any given order in the derivatives) from these asymptotic tensor structures. At lowest order, i.e.,  no derivatives \emph{and} flat intrinsic geometry, the correspondence is trivial as it is non-dynamical. Indeed, computing the asymptotic stress tensor and current of the solution \eqref{solutionSCH}, one obtains the stress tensor and current \cite{Emparan:2011hg}
\begin{equation}\label{perstess}
T_{ab}=\varrho \thinspace u_a u_b + P \Delta_{ab} \ , \quad j=Q\star 1 \ .
\end{equation}
Here $\varrho$, $P$ and $Q$ denotes the energy (density), pressure and charge of the brane and can be computed from the Gibbs free energy $G$, which in turn is computed from the on-shell action and takes the form
\begin{equation}\label{gibbsfree}
G=\text{Vol}(S^{n+1})\left(\frac{n}{4\pi \thinspace \mathcal T}\thinspace \sqrt{\left(1- \frac{\Phi^2}{N} \right)^{\thinspace N}}\right)^n \ .
\end{equation}
The temperature $\mathcal T$ and the chemical potential $\Phi$ are the intensive thermodynamic variables which do not depend on Newton's constant and are easily written in terms of the parameters $r_0$ and $\gamma_0$ using standard methods, 
\begin{equation}\label{r0gamma0}
\mathcal T=\frac{n}{4\pi r_0\sqrt{\left(1+\gamma_0\right)^N}} \ , \quad \Phi=\sqrt{\frac{N\gamma_0}{1+\gamma_0}} \ .
\end{equation}
Notice that the entropy $s$ and charge $Q$ are conjugate to $\mathcal T$ and $\Phi$ and are computed from $G$ in the usual way. The energy density $\varrho$ and pressure $P$ are then derived using the Gibbs-Duhem relation $w \equiv \varrho+P=\mathcal T s$ and the defining identity $G=-P-\Phi Q$. At lowest order and flat intrinsic geometry, the correspondence between fluid dynamics and gravity is therefore just a convenient repackaging of black hole thermodynamics in terms of a relativistic (perfect) fluid. However, note that if one abandons the requirement of flat intrinsic geometry, the statement becomes an equivalence between gravity and perfect fluid dynamics on a curved $p$-submanifold (known as \emph{the blackfold approach}, see \cite{Emparan:2011br} for a review), which is a non-trivial statement. In this work we keep the intrinsic geometry flat, or equivalently, we do not perturb the transverse sphere $S^{n+1}$. 

\subsection{The perturbative expansion} \label{outlinepertexp}

In order to carry out the perturbative procedure we proceed in the standard way. Here we will give a brief summary of the computation and refer to Appendix~\ref{app:details} for many of the details (also see the papers \citep{Bhattacharyya:2008jc, Camps:2010br, Gath:2013qya}). In order to ensure that the perturbative problem is well-posed, we need to cast the metric \eqref{solutionSCH} into Eddington-Finkelstein (EF) form (i.e., a coordinate transformation $x^a\to \sigma^a(r)$ tailored so that $|\td r|=0$). In these coordinates the metric \eqref{solutionSCH} takes the form
\begin{equation}\label{worldvolgeometry}
\td s^2=h^{-\frac{Nn}{p+n+1}}\Bigg(\Big( - f u_a u_b  +\Delta_{ab} \Big) \td \sigma^a \td \sigma^b - 2 h^{\frac{N}{2}}u_a \thinspace \td \sigma^a \td r + h^N r^2 \thinspace \td \Omega_{(n+1)}^2 \Bigg) \ .
\end{equation}
Transforming the coordinates of course also induces a transformation of the gauge field, however, one can easily show that in the case of Schwarzschild $\to$ EF coordinates the transformation of the gauge field can be undone by a suitable gauge transformation. We therefore keep the form of the gauge field \eqref{solutionSCH} consistent with our gauge choice introduced below. According to the ideas of fluid/gravity outlined above, the solution \eqref{worldvolgeometry} (along with the associated matter fields) is part of a larger class of solutions $\td s_f^2$ for which the parameters $u^a$, $r_0$ and $\gamma_0$ (collectively denoted by $\xi=(u^a,r_0,\gamma_0)$) are worldvolume fluctuating functions. This solution, which reduces to \eqref{worldvolgeometry} for constant $u^a$, $r_0$ and $\gamma_0$, is in general unknown, but can be probed perturbatively in a derivative expansion. Therefore for long-wavelength perturbations, 
\begin{equation} \label{fullsol}
\td s_f^2=\td s^2 + \td s_\partial^2 + \mathcal{O}(\partial^2) \ , \quad A_f=A + A_{\partial} + \mathcal{O}(\partial^2)  , \quad \phi_f = \phi + \phi_\partial + \mathcal O (\partial^2) \ .
\end{equation}
Here $\td s^2$ is the geometry \eqref{worldvolgeometry} expanded to first order in worldvolume derivatives and $\td s_\partial^2$ denotes the first-order correction coming from the full solution $\td s_f^2$ (and similarly for the matter fields). The main purpose of this section is to compute the first-order expansion of the full solution and extract the first-order effective currents. The EOMs exhibit a large gauge redundancy. In order to simplify the computations it is convenient to choose a (consistent) gauge where the transverse components of $(A_\partial)$ are taken to zero and furthermore $(g_\partial)_{rr}=0$, $(g_\partial)_{\Omega\Omega}=0$. The latter gauge choice allows us to reduce over the transverse sphere $S^{n+1}$, effectively leaving us with a $p+2$ dimensional problem. Notice that although the transverse $S^{n+1}$ drops out of the problem, the parameter $n$ still plays an important r\^{o}le as it will enter the various coefficients in the resulting set of equations. Choosing the appropriate ansatz for the perturbations (collectively denoted by $\psi_\partial=(g_\partial,A_\partial,\phi_\partial)$) and plugging them into the EOMs \eqref{EOMsip} produces two sets of qualitatively different equations, schematically of the form
\begin{equation}
\textit{Constraint:} \ \ \mathbb C_\partial \ \partial \xi  + \mathcal O (\partial^2) = 0 \ , \quad \quad \textit{Dynamical:}  \ \ \mathbb L^{(1)}_{r} \mathbb L^{(2)}_{r} \  \psi_\partial = s_\partial(r) + \mathcal O (\partial^2) \ .
\end{equation}
Here $\mathbb C_\partial$ is an operator acting on the derivatives of the intrinsic fields $\xi$ and $L^{(1)}_{r}$ and $L^{(2)}_{r}$ are two first-order linear differential operators acting on the perturbations $\psi_\partial$ as a function of the radial coordinate $r$ (and only $r$) and finally $s_\partial(r)$ is a source term (which is a rational function in $r$ whose coefficients depend on $\partial \xi$). We note that in order to obtain the dynamical equations one has to explicitly use the constraint equations.  The constraint equations are independent of the radial coordinate and, as the name suggests, put constraints on (i.e., relations between) the derivatives of the intrinsic fields. As expected, the constraint equations are found to be equivalent to the conservation equations of the leading order perfect fluid stress tensor and current \eqref{perstess}. Notice that the conservation of $j$ takes the form $\partial_a Q=0$, which just expresses that the charge $Q$ is non-fluctuating along the worldvolume. Locally the inverses of the differential operators $\mathbb L^{(1)}_{r}$ and $\mathbb L^{(2)}_{r}$ exist and the formal solution to the dynamical equations reads
\begin{equation}\label{solgen}
\psi_\partial=\left( \mathbb L^{(2)}_{r} \right)^{-1}\left( \mathbb L^{(1)}_{r} \right)^{-1}s_\partial + \left( \mathbb L^{(2)}_{r} \right)^{-1} h_1 + h_2 \ .
\end{equation}
Here $h_1$ and $h_2$ denote functions in the kernel of $\mathbb L^{(1)}_{r}$ and $\mathbb L^{(2)}_{r}$, respectively. We note that the inverse operators of $\mathbb L^{(1)}_{r}$ and $\mathbb L^{(2)}_{r}$ in general involve various integrations which when evaluated on $s_\partial$ are quite complicated leading to various types of hypergeometric functions. The local decomposition \eqref{solgen} is of course not unique, however, if we require the inverse of $\mathbb L^{(1)}_{r}$ to exist globally (in particular at $r=r_0$) or equivalently require horizon regularity, the homogeneous solution $h_1$ is forced to vanish leaving only $h_2$, which is then fixed by the boundary conditions and choice of fluid frame. Indeed, a subset of the freedom in the homogeneous solution $h_2$ comes directly from the seed solution \eqref{worldvolgeometry} and can be generated by $\mathcal O (\partial)$ shifts $r_0\to r_0 + \delta r_0$, $\gamma_0\to \gamma_0 + \delta \gamma_0$ and $u^a \to u^a + \delta u^a$ (along with gauge transformations of $A_f$). Such shifts of course map to new (regular) asymptotically flat solutions which differ from \eqref{worldvolgeometry} by $\mathcal O(\partial)$. From a relativistic fluid point of view, this freedom is expected and corresponds to $\mathcal O (\partial)$ redefinitions of the temperature, chemical potential and fluid velocity. In order to fix the gauge (the fluid frame), we require that the first-order solution reduces to \eqref{worldvolgeometry} when the fluctuations vanish, which in turn corresponds to choosing the Landau frame on the fluid side. The remaining freedom in $h_2$ parameterizes non-renormalizable modes which we require to vanish by virtue of asymptotic flatness.\footnote{For $n=1$ one needs to fix some additional gauge freedom in $g_{\partial ri}$ in order to ensure asymptotically flatness.} In this way the freedom in the homogeneous solution $h_2$ can be completely removed and we obtain the full solution to first order in the derivative expansion. 

\subsection{Transport coefficients} \label{sec:qptransport}

Having obtained the first-order regular asymtotically flat corrected solution and reexpressing it in Schwarzschild coordinates, it is straightforward to compute the induced effective stress tensor using familiar techniques and extract the first-order transport coefficients. By direct computation we obtain the following effective stress tensor
\begin{equation} \label{eqn:firstorderfluid}
T_{ab}=\varrho \thinspace u_a u_b + P \Delta_{ab}  + \Pi_{ab}^{(1)} + \mathcal{O}(\partial^2) \quad \text{with} \quad \Pi_{ab}^{(1)}= - 2 \eta \sigma_{ab} - \zeta \vartheta \Delta_{ab} \ . 
\end{equation} 
Here $\sigma_{ab}$ and $\vartheta$ are the (fluid) shear and expansion of the congruences $u^a$, respectively, and the coefficients $\eta$ and $\zeta$ are the corresponding shear and bulk viscosity.\footnote{Notice that for $p=1$ the shear tensor vanishes.} They explicitly evaluate to 
\begin{equation}\label{transporttopform}
\eta=\frac{s}{4\pi} \ , \quad \frac{ \zeta }{ \eta} = 2\left(\frac{1}{p}+\frac{C-2n}{n+1+C\gamma_0}\thinspace \gamma_0 + \frac{(n+1)\big(1+(C-2n)\gamma_0\big)}{\left(n+1+C\gamma_0\right)^2}\right) ~~,
\end{equation} 
where we have defined the constant $C \equiv 2-n(N-2)$. For fixed $r_0$, the viscosities are parametrized by the charge parameter $\gamma_0$ and the parameter $N$. The neutral limit can be obtained independently by taking either of the parameters to zero. Indeed, for large values of the dilaton coupling the dynamics of the brane effectively reduces to that of the neutral black $p$-brane (we have checked that our results reduce to those obtained in the neutral limit \cite{Camps:2010br}). We note that the shear viscosity increases with the charge while bulk-to-shear viscosity ratio decreases as charge is added to the brane (for fixed $r_0$). As expected the gravitational solution saturates the hydrodynamic bound $\eta/s\geq 1/4\pi$ \citep{Kovtun:2003wp, *Buchel:2003tz}. We also note that the charge current $j= Q \star 1$ does not receive any first-order contribution. This is of course just a reflection of current conservation in the effective theory. Finally, we have checked that our results agree with those obtained for the D3-brane in Ref.~\citep{Emparan:2013ila} (here $p=3$, $D=n+p+3=10$, $a=0$).\footnote{One has to employ the limit $R_c\to 1$ and make use of the identifications $r_-^4=r_0^4\gamma_0$, $r_+^4=r_0^4(1+\gamma_0)$ and $\gamma_0=(1-\delta_e)/\delta_e$.}



\subsection{Dynamical stability} \label{sec:stability}
\label{sec:qpstability}


With the fluid transport determined, we can now address the stability properties of fundamentally charged branes by analyzing the response of their corresponding effective fluids under small long-wavelength perturbations. The dynamics of a charged fluid is governed by the worldvolume conservation equations given by Eq.~\eqref{eqn:conservation}. Here, we are interested in the case where the stress tensor $T$ takes the form of a general dissipative fluid to first order \eqref{eqn:firstorderfluid}. For the sound mode(s) we write the dispersion relation, valid up to second order in the wave vector $k = \sqrt{k_ik^i}$,\footnote{Here the relativistic wave vector is $k^{a} = \{\omega, k^i\}$.} as
\begin{equation}\label{eqn:sosmode}
\omega(k)=\pm c_s k +  i \mathfrak{a}_s k^2 ~~.
\end{equation}
Here $c_s$ is the speed of sound given by $c_s^2 = \nicefrac{\partial P}{\partial \varrho}$, where the thermodynamical quantity kept fixed when taking the derivative depends on the specific type of charged fluid in question. The attenuation coefficient $\mathfrak{a}_s$ controls the dampening of the longitudinal sound mode and is determined by the first-order transport coefficients \eqref{transporttopform}.
In order for the fluid to be dynamically stable we must require that the speed of sound squared $c_s^2$ and the attenuation coefficient $\mathfrak{a}_s$ are both positive. In addition to the sound mode, the fluid also exhibits a transverse shear mode,
\begin{equation} \label{eqn:shearmode} 
\omega(k)=\frac{i\eta}{w}k^2 ~~,
\end{equation}
which is stable provided that $\eta > 0$. 

\begin{figure}[t]
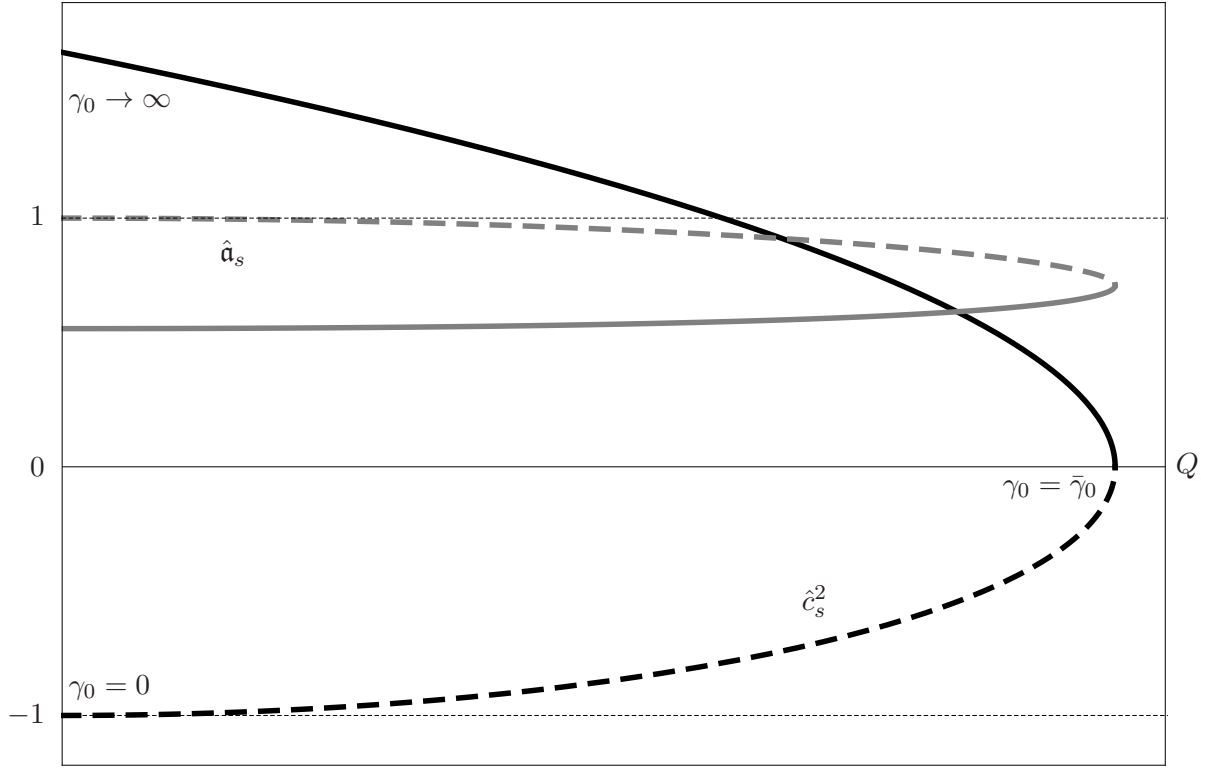

\begin{center}
\begin{lpic}{12232014qpsoundmode(0.45,)}
\lbl[r]{-5,88;$0$}
\lbl[r]{-5,162;$1$}
\lbl[r]{-5,15;$-1$}
\lbl[l]{2,23;$\gamma_0 = 0$}
\lbl[l]{275,82;$\gamma_0 = \bar{\gamma}_0$}
\lbl[l]{2,195;$\gamma_0 \rightarrow \infty$}
\lbl[t]{329,92;$Q$}
\lbl[t]{220,53;$\hat{c}^2_s$}
\lbl[t]{50,155;$\hat{\mathfrak{a}}_s$}
\end{lpic}
\caption{The qualitative behavior of the coefficients of the sound mode given by Eq.~\eqref{eqn:sosmode} as a function of the charge $Q$ for a fixed temperature $\mathcal{T}$. The quantities $\hat{c}^2_s$ and $\hat{\mathfrak{a}}_s$ correspond to the speed of sound squared and the attenuation coefficient normalized with respect to their neutral values, respectively. There is one unstable branch connected to the neutral limit ($\gamma_0 \rightarrow 0$) plotted with dashed lines and one stable branch connected to the extremal limit where both coefficients are positive. The stable branch is reached exactly when the charge parameter $\gamma_0$ exceeds the threshold $\bar{\gamma}_0$ given by Eq.~\eqref{eqn:qpthreshold}. This point corresponds to the maximal charge of the brane configuration for a given fixed temperature.}
\label{fig:qpsoundmode}
\end{center}
\end{figure}

Now, for fundamentally charged $p$-branes the charge is not allowed to redistribute itself on the worldvolume of the brane, since it is conserved along all directions as noted previously. This means that the dynamics of the effective fluid will be reminiscent of that of a neutral brane, since the charge only plays a r\^{o}le in the equation of state. In particular, the leading order stability is solely determined by the sound mode \eqref{eqn:sosmode}. 
The system will therefore be stable to linear order in $k$ as long as the speed of sound squared,
\begin{align} \label{eqn:qpsos}
	c_s^2 =& \left(\frac{\partial P}{\partial \varrho}\right)_{Q_p} = -\frac{1 + (2-nN)\gamma_0}{n+1 + C\gamma_0} ~~,
\end{align}
is positive. We note that in the limit of vanishing charge ($\gamma_0 \rightarrow 0$) we recover $c_s^2=\nicefrac{-1}{(n+1)}$ \citep{Camps:2010br} signifying that the neutral brane exhibits a Gregory-Laflamme (GL) instability \citep{Gregory:1993vy, *Gregory:1994bj}.
However, for finite $\gamma_0$ we find, as in Ref.~\cite{Emparan:2011hg}, that there exists a threshold,
\begin{equation} \label{eqn:qpthreshold}
	\bar{\gamma}_0 =  \frac{1}{nN - 2} ~~,
\end{equation}
above which the black brane is stable under long-wavelength perturbations to linear order. 
For fixed temperature, the threshold value $\bar{\gamma}_0$ is precisely where the brane configuration reaches its maximal charge. As illustrated in Fig.~\ref{fig:qpsoundmode}, this point exactly corresponds to a transition point between an unstable branch and a stable branch of configurations with the neutral brane configuration located at the endpoint $\gamma_0 \to 0$ of the unstable branch. For sufficiently large dilaton coupling all fundamentally charged $p$-branes are therefore unstable, since they approximates the neutral brane.
On the other hand, the stable branch connects with the extremal limit and since $\gamma_0$ measures the ratio between local electrostatic energy and thermal energy of the fluid, the stable regime is where the electrostatic energy is dominant.
Notice that the limit $\gamma_0 \to \infty$ corresponds to flat space.
However, it is not entirely clear if any of the hydrodynamic interpretation survives in the strict limit.
The threshold value does not exist in instances where $nN < 2$, since $\gamma_0$ is a non-negative parameter. In those cases, the speed of sound is imaginary for all values of $\gamma_0$.
Finally, for the branes in ten and eleven dimensional supergravity we have listed several values of interest in Table~\ref{tab:g0critvales}. In particular, we note that the threshold $\bar{\gamma}_0$ does not exist for $p=5,6$ in $D=10$. This is in agreement with the expectation that the supergravity descriptions of D/NS/M-branes are stable with the exception of the D5, D6 and NS5 brane \cite{Reall:2001ag, Gregory:1993vy, Gregory:1994bj, Emparan:2011hg}.


\begin{table}[t]
\begin{center}
\begin{tabular}{@{}lllll@{}}
  \toprule
	Brane & $a$ & $\bar{\gamma}_0$ & $c_s^2$ & $\nicefrac{w \mathfrak{a}_s}{\eta}$\\
	\midrule
	D$p$ & $\frac{3-p}{2}$ & $\frac{1}{5-p}$ & $\frac{(5-p) \gamma_0 - 1}{8-p + (9-p) \gamma_0}$ & 
	$\frac{72-17p+p^2+8(8-p)\gamma_0+4(9-p)\gamma_0^2}{(8-p+(9-p)\gamma_0)^2}$
	\\[4mm]
	M2 & 0 & $\frac{1}{4}$ & $\frac{4\gamma_0 - 1}{7 + 8\gamma_0}$ & $\frac{1}{2}\left(1 + \frac{63}{(7+8\gamma_0)^2}\right)$ \\[4mm]
	M5 & 0 & $1$ & $\frac{\gamma_0 - 1}{4 + 5\gamma_0}$ & $ \frac{4}{5} \left(1 + \frac{9}{(4+5\gamma_0)^2}\right)$ \\[1.5mm]
	\bottomrule
\end{tabular}
	\caption{The expansion coefficients of the sound mode \eqref{eqn:sosmode} for the $p$-branes of ten and eleven dimensional supergravity ($N=1$). For a sufficiently large value of $\gamma_0$ the threshold value $\bar{\gamma}_0$ is exceeded and the fundamentally charged branes are stable. We note that the threshold value in ten dimensional supergravity increases with the spatial dimension and that black NS/D$p$-branes are always unstable for $p \geq 5$. It is also worth noting that the values for the D1 and D4 are equivalent to those of the M2 and M5, respectively, due to Type II A $\leftrightarrow$ M-theory relation. Finally, the values for the fundamental string and NS5 are equivalent to those of the D1 and D5, respectively, as $N$ is invariant under $a \rightarrow -a$.}
	\label{tab:g0critvales}
\end{center}
\end{table}


We can now proceed by refining the analysis to quadratic order in $k$ by including the second-order term in the sound mode. As explained, in the absence of charge diffusion, the attenuation coefficient will take the exact same form as a neutral fluid, hence
\begin{equation} \label{eqn:qpsoundmodeatt}
	\mathfrak{a}_s = \frac{1}{w} \left(\left(1-\frac{1}{p}\right)\eta+\frac{\zeta}{2}\right) ~~.
\end{equation}
The effects of the fluid carrying charge therefore only appears through the explicit dependence on $Q$ in the shear and bulk viscosities given by Eq.~\eqref{transporttopform}. When the threshold \eqref{eqn:qpthreshold} exists, we find, as shown in Fig.~\ref{fig:qpsoundmode}, that the attenuation coefficient is positive for both branches of configurations. The stability is therefore fully determined by the linear order, i.e., by the speed of sound. This is in contrast to configurations with smeared charges where the attenuation coefficient plays an important r\^{o}le for the stability properties of the effective fluid as we shall see in Sec.~\ref{sec:q0stability}. When the threshold exists, the fundamentally charged $p$-brane is therefore dynamically stable for sufficiently large charge parameter $\gamma_0$ at least to next-to-leading order.

An occurrence of a dynamical GL-like instability is conjectured to be intercorrelated with a thermodynamical stability \citep{Gubser:2000ec, *Gubser:2000mm, Reall:2001ag} \textit{viz.} the correlated stability conjecture. It is therefore interesting to relate the above dynamical analysis to the thermodynamic properties of the effective fluid. The condition for thermodynamic stability is computed in the canonical ensemble, since the charge is fixed in the system and thus only requires positivity of the specific heat $c_{Q_p}$. Using the thermodynamic quantities \eqref{r0gamma0} and the expression for the speed of sound \eqref{eqn:qpsos} one can show that the point where the specific heat becomes positive overlaps precisely with the threshold value given by Eq.~\eqref{eqn:qpthreshold}, i.e., the configuration with the maximally allowed charge (for a given temperature). Indeed, as pointed out in Ref.~\cite{Emparan:2011hg}, there is a direct relation between the speed of sound and the specific heat $c_{Q_p}$ at fixed charge given by
\begin{equation}
	c_s^2 = \left(\frac{\partial P}{\partial \varrho}\right)_{Q_p} = s \left( \frac{\partial \mathcal{T}}{\partial \varrho} \right)_{Q_p} = \frac{s}{c_{Q_p}}  ~~.
\end{equation}
For this system, we therefore find that the dynamical stability is in agreement with the correlated stability conjecture. Although, our results are only valid to first order we expect the statement to hold to all orders.



\section{Hydrodynamics of dilatonic Maxwell charged branes}\label{pequal0hydro}

In this section, we write down the first-order effective hydrodynamic expansion for the $p$-branes of the theory \eqref{action} with $q=0$ (Einstein-Maxwell-Dilaton (EMD) theory). These results provide the dilaton generalization of the results originally presented in Ref. \cite{Gath:2013qya}. In terms of the parameter $N$, the dilaton coupling $a$ now takes the form, 
\begin{equation}
a^2=\frac{4}{N}-2\left(\frac{D-3}{D-2}\right) \ .
\end{equation}
In particular, for the parameters relevant for type IIA supergravity, we have $a=3/2$, and the action \eqref{action} is that appropriate for describing smeared configurations of D$0$-branes. The leading order solution is given by
\begin{equation}\label{solutionSCHMaxwell}
\begin{split}
\td s^2&=h^{-\left(\frac{n+p}{n+p+1}\right)N}\Bigg(-f u_a u_b \thinspace \td x^a \td x^b + h^N\left(\Delta_{ab} \thinspace\td x^a \td x^b+f^{-1}\td r^2+r^2 \thinspace \td \Omega_{(n+1)}^2\right) \Bigg) \ , \\ 
&\ \ \quad \phi=\frac{aN}{4}\log h \ , \quad \quad A_{(1)}= \sqrt{N\left(\frac{\gamma_0+1}{\gamma_0}\right)} \thinspace \left(h^{-1}-1\right) u_a \td x^a \ .
\end{split}
\end{equation} 
The effective stress tensor and Maxwell current are computed from the asymptotics of the solution and take the perfect fluid form, 
\begin{equation}\label{perstessmaxwell}
T_{ab}=\varrho \thinspace u_a u_b + P \Delta_{ab} \ , \quad j_a=\mathcal Q \thinspace u_a \ .
\end{equation}
Here the energy density $\varrho$, pressure $P$ and charge density $\mathcal Q$ are obtained from the free energy in the usual way. Direct computation of the on-shell action reveals that the free energy $G$ again takes the from \eqref{gibbsfree}. Moreover, the two parameters $r_0$ and $\gamma_0$ are related to the temperature $\mathcal T$ and chemical potential $\Phi$ as in Eq. \eqref{r0gamma0}. However, the Gibbs-Duhem relation now takes the form $w \equiv \varrho+P=\mathcal T s + \Phi \mathcal Q$. Here we have used a calligraphed $\mathcal Q$ to denote the monopole charge density in order to distinguish it from the fundamental dipole type charge $Q$ considered in the previous section.

\subsection{Transport coefficients} \label{sec:q0transport}

Carrying out the perturbative computation of the first-order corrected fields and the corresponding effective currents follows the procedure explained in Sec.~\ref{pequalqhydro} (see Appendix~\ref{app:details} for many of the details and e.g. Ref.~\cite{Gath:2013qya}). The most important difference between the two computations consists of the existence of an additional $\text{SO}(p)$ vector (dynamical) equation in the Maxwell case. This is in accordance with the fact that a worldvolume one-form potential contains $p$ more degrees of freedom than a top-form potential. The constraint equations coming from Einstein's equations take the same form as before but the constraint equation deriving from the Maxwell equation is shown to be equivalent to current conservation, $\partial_a j^a=0$. This allows for fluid dynamical fluctuations in the charge density $\mathcal Q$, which in turn shows up as a new charge diffusion transport coefficient in the effective theory. The most general first-order derivative corrected effective current, consistent with the second law of thermodynamics, takes the form (here written in the Landau frame)
\begin{equation}\label{effcurrent}
j^a=\mathcal Q u^a +\Upsilon^a_{(1)} + \mathcal{O}(\partial^2) \quad \text{with} \quad 
\Upsilon^a_{(1)}=-\mathfrak D \left(\frac{\mathcal Q \mathcal T}{w}\right)^2\Delta^{ab}\partial_a\left(\frac{\Phi}{\mathcal T}\right) \ .
\end{equation}
Here $\mathfrak D$ is the transport coefficient associated with diffusion (appropriately normalized). 

Without further ado, we now present our results for the transport coefficients. The shear and bulk viscosities are given by
\begin{equation} \label{eqn:q0shearbulk}
\eta=\frac{s}{4\pi} \ , \quad \frac{\zeta}{\eta}=\frac{2}{p}+\frac{2}{C}\left(2-N+\frac{(n+1)N}{(n+1+C\gamma_0)^2}\right) \ ,
\end{equation}
here $C$ is the constant introduced below Eq. \eqref{transporttopform}. 
Again, for fixed $r_0$, the viscosities are parametrized by $\gamma_0$ and $N$ and they reduce to the values of the neutral limit if either of the parameters are taken to zero.
As expected, the shear viscosity saturates the bound $\eta/s\geq 1/4\pi$. The diffusion constant $\mathfrak D$, associated with the effective current (Eq. \eqref{effcurrent}), is determined to 
\begin{equation} \label{eqn:q0diffusion}
\frac{\mathfrak{D}}{\eta} = \frac{4\pi r_0 (1+\gamma_0)}{nN \gamma_0 \sqrt{(1+\gamma_0)^N}}   ~~,
\end{equation}
As is straightforward to check, these results agree with those of \cite{Gath:2013qya} in the limit where the dilaton coupling goes to zero, or equivalently, $N\to \frac{2(D-2)}{D-3}$.

\subsection{Dynamical stability} \label{sec:q0stability}




We now analyze the dispersion relations of the fluid associated to the Maxwell charged branes. In contrast to the fundamentally charged branes, discussed previously, the Maxwell charged branes exhibit charge diffusion which significantly changes the effective dynamics. Most notably, since the charge density can redistribute over the worldvolume, this gives rise to an additional longitudinal mode with the following dispersion relation,
\begin{equation} \label{eqn:diffmode}
	\omega(k) = i \mathfrak{a}_\mathfrak{D} k^2 ~~.
\end{equation}
This mode is associated to the attenuation of the (long-wavelength) diffusion of charge and is a first-order derivative effect. The existence of this charge diffusion mode explicitly means that the stability of the system to second order in $k$ is not only determined by the attenuation coefficient $\mathfrak{a}_s$ of the sound mode \eqref{eqn:sosmode}, but also by the attenuation coefficient $\mathfrak{a}_\mathfrak{D}$. In order for the system to be stable to second order one must therefore require both to be positive.
As before this system also exhibits a transverse shear mode given by Eq.~\eqref{eqn:shearmode}. However, since the shear viscosity is positive, this mode does not play any r\^{o}le for the stability properties of the fluid.



\begin{figure}[t]
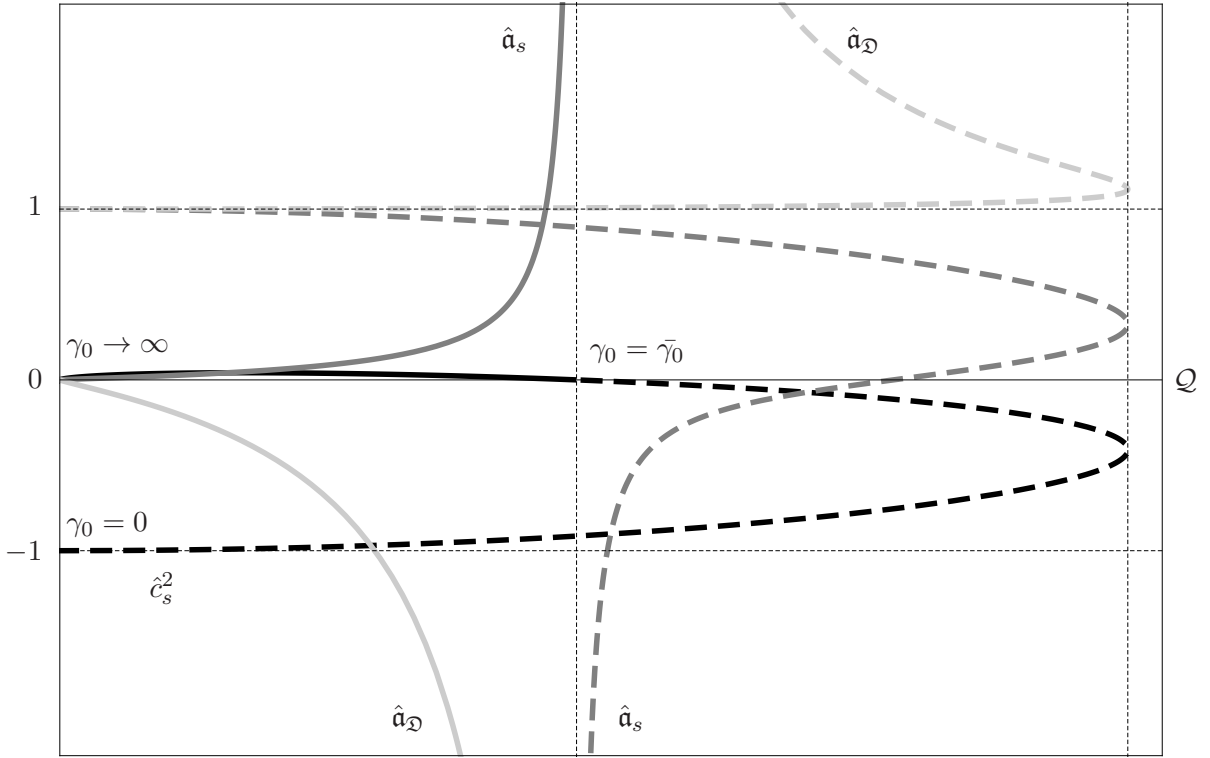

\begin{center}
\begin{lpic}{12232014q0modes(0.45,)}
\lbl[r]{-5,162;$1$}
\lbl[r]{-5,111;$0$}
\lbl[r]{-5,60;$-1$}
\lbl[l]{2,68;$\gamma_0 = 0$}
\lbl[l]{155,119;$\gamma_0 = \bar{\gamma_0}$}
\lbl[l]{2,120;$\gamma_0 \rightarrow \infty$}
\lbl[t]{329,114;$\mathcal{Q}$}
\lbl[t]{30,54;$\hat{c}_s^2$}
\lbl[t]{167,15;$\hat{\mathfrak{a}}_s$}
\lbl[t]{133,215;$\hat{\mathfrak{a}}_s$}
\lbl[t]{102,15;$\hat{\mathfrak{a}}_{\mathfrak{D}}$}
\lbl[t]{235,215;$\hat{\mathfrak{a}}_{\mathfrak{D}}$}
\end{lpic}
\caption{The qualitative behavior of the coefficients of the longitudinal modes given by Eq.~\eqref{eqn:sosmode} and Eq.~\eqref{eqn:diffmode} as a function of the charge density $\mathcal{Q}$ for fixed local temperature $\mathcal{T}$. The quantities $\hat{c}^2_s$, $\hat{\mathfrak{a}}_s$ and $\hat{\mathfrak{a}}_{\mathfrak{D}}$ are given by the Eqs.~\eqref{eqn:q0sos}, \eqref{eqn:q0sosatt} and \eqref{eqn:q0diffatt} normalized with respect to their individual neutral values. 
Dashed lines are for $\gamma_0 < \bar{\gamma_0}$ while filled lines are for $\gamma_0 > \bar{\gamma_0}$.
We observe that no region exists where all three quantities are positive at the same charge density $\mathcal{Q}$.
Furthermore, the attenuation coefficients both show a hyperbolic behavior at the threshold $\bar{\gamma_0}$ indicated by a vertical dashed line.
Note that due to the choice of normalization of the diffusion coefficient in Eq.~\eqref{effcurrent}, the attenuation coefficient $\bar{\mathfrak{a}}_{\mathfrak{D}}$ takes a fictitious finite value in the neutral limit.
%
}
\label{fig:q0soundmode}
\end{center}
\end{figure}

To linear order in $k$, the stability of the Maxwell system is, similarly to the fundamentally charged brane, solely determined by the sound mode. More precisely, it is dictated by the speed of sound squared which for the Maxwell system is given by
\begin{equation} \label{eqn:q0sos}
	c_s^2 = \left(\frac{\partial P}{\partial \varrho}\right)_{\tfrac{s}{\mathcal{Q}}} = -\frac{1 + (2-N)\gamma_0}{(1 + \gamma_0 N) (n+1 + C\gamma_0)} ~~.
\end{equation}
In the limit of vanishing charge ($\gamma_0 \to 0$) we again recover the neutral value for the speed of sound. For finite $\gamma_0$ we find the threshold,
\begin{equation}\label{eqn:q0stabcond}
	\bar{\gamma}_0 = \frac{1}{N-2} ~~,
\end{equation} 
above which the Maxwell charged black brane is stable to leading order. 
Note that the threshold only exists for $N>2$, or equivalently, when the dilaton coupling $a$ is sufficiently small.\footnote{Explicitly the dilaton coupling squared has to be smaller than $\bar{a}^2 = \frac{2}{n+p+1}$.}
The qualitative behavior of the speed of sound (for $N>2$) is illustrated in Fig.~\ref{fig:q0soundmode}. In particular, we find that there exist two branches of brane configurations for the same charge density; an unstable branch ($\gamma_0 < \bar{\gamma}_0$) connected to the neutral brane configuration and a stable branch ($\gamma_0 > \bar{\gamma}_0$) connected to the extremal brane configuration.
However, in contrast to the fundamentally charged brane, the merging point (defined by $\bar{\gamma}_0$) does not coincide with the maximal charge configuration.
On the other hand, if the threshold does not exist, i.e., $N<2$, the speed of sound is imaginary for all charge densities. The system is thus unstable independent of the charge.
This instability is in accordance with the one observed for configurations of smeared D0-branes ($N=1$) \citep{Bostock:2004mg, *Aharony:2004ig}. In Table~\ref{tab:q0table}, we list the values of interest connected to this special case.

\begin{table}[t]
\begin{center}
\begin{tabular}{@{}lllllll@{}}
	\toprule
	Brane & $a$ & $\bar{\gamma}_0$ & $c_s^2$ & $\nicefrac{w\mathfrak{a}_s}{\eta}$ &
	$\nicefrac{w\mathfrak{a}_{\mathfrak{D}}}{\eta}$ \\
	\midrule	
	D$0$ & $\frac{3}{2}$ & $-1$ & $-\frac{1}{8-p+(9-p)\gamma_0}$ &
	$(9-p)(8-p)(\gamma_0+1)^2c_s^4$ &
	$1$ \\[1mm]
	\bottomrule
\end{tabular}
	\caption{Coefficients of the longitudinal sound and diffusion mode for the black D0-brane smeared in $p\geq1$ directions. Note that the first-order coefficient (speed of sound squared) is negative whereas both of the second-order attenuation coefficients are positive. 
	}
	\label{tab:q0table}
\end{center}
\end{table}








We now address the stability to quadratic order in $k$. We therefore consider the two attenuation coefficients $\mathfrak{a}_s$ and $\mathfrak{a}_\mathfrak{D}$ associated to the sound and diffusion mode, respectively.
For the sound mode we find, in contrast to Eq.~\eqref{eqn:qpsoundmodeatt}, the following modification due to the presence of charge diffusion,
\begin{equation} \label{eqn:q0sosatt}
	\mathfrak{a}_s = \frac{1}{w} \left(\left(1-\frac{1}{p}\right)\eta+\frac{\zeta}{2} + \frac{2}{\mathcal{T}} \frac{\mathcal{Q}^2}{c_s^2} \left( \frac{\mathcal{Q}^2}{w \Phi} \thinspace \frac{c}{C_{\mathcal{Q}}} \right)^2 \mathfrak{D} \right)  ~~,
\end{equation}
where we have introduced the specific heat capacity $C_{\mathcal{Q}}$ and the (inverse) isothermal permittivity $c$ given in Eq.~\eqref{eqn:q0specificheat}. 
For the diffusion mode \eqref{eqn:diffmode}, we find 
\begin{equation} \label{eqn:q0diffatt}
	\mathfrak{a}_\mathfrak{D} = \frac{\mathcal{Q}^2}{c_s^2 w} \thinspace \frac{c}{C_{\mathcal{Q}}} \mathfrak{D} ~~.
\end{equation}
Assuming that $N>2$, the attenuation coefficients exhibit a hyperbolic divergence around $\bar{\gamma}_0$.
We refer to Fig.~\ref{fig:q0soundmode}, where we have plotted a generic case of the two longitudinal modes.
This divergent behavior at $\bar{\gamma}_0$ is very different compared to the continuous behavior observed for $\mathfrak{a}_s$ in Sec.~\ref{sec:qpstability} (see Fig.~\ref{fig:qpsoundmode}). 
Indeed, it seems that the linearized analysis breaks down at the threshold $\bar{\gamma}_0$.
However, besides this new feature we still find that both the speed of sound squared and the attenuation coefficient $\mathfrak{a}_s$ are positive when $\gamma_0$ surpasses the threshold. The stability of the sound mode is therefore, similarly to the fundamentally charged branes, solely dictated by the speed of sound.
In contrast to the attenuation coefficient of the sound mode, it is quite apparent from Eq.~\eqref{eqn:q0diffatt}, that the hyperbolic feature of the diffusion attenuation coefficient is simply dictated by the inverse speed of sound.
We therefore find that the diffusion mode has exactly the complementary behavior of the sound mode, that is, when the sound mode is stable the charge diffusion mode is unstable and vice versa. 
The Maxwell charged brane configurations therefore suffer a GL instability for all charge densities.



Finally, we check whether the dynamical GL-like instability observed above is connected with the thermodynamic stability of the system as predicted by the correlated stability conjecture.
The conditions for thermodynamic stability of the Maxwell black brane are computed in the grand canonical ensemble since charge is allowed to redistribute itself in the directions of the brane. This exactly leads to the requirement of positive specific heat capacity and positive (inverse) isothermal permittivity \citep{Ross:2005vh, Harmark:2005jk},
\begin{equation} \label{eqn:q0specificheat}
\begin{split}
  C_{\mathcal{Q}} = \left(\frac{\partial\varrho}{\partial\mathcal{T}} \right)_{\mathcal{Q}} &= \left( \frac{n + 1 + C\gamma_0}{(nN - 2)\gamma_0-1} \right) s  ~~, \\
  c = \left( \frac{\partial\Phi}{\partial\mathcal{Q}} \right)_{\mathcal{T}} &= \left( \frac{1}{(\gamma_0+1) (1 -  (nN-2)\gamma_0)} \right) \frac{1}{s\mathcal{T}}  ~~.
\end{split}
\end{equation}
It is straightforward to see that these two conditions $C_{\mathcal{Q}} > 0$ and $c > 0$ are complementary and can never be satisfied. Indeed, the two quantities exchange signs at $\gamma_0 = \nicefrac{1}{(nN-2)}$. 
Comparing with the dynamical analysis above, we observe that it is not sufficient to consider the dispersion relations to leading order, i.e., only considering the speed of sound $c_s$, since it predicts the system to be stable above the threshold \eqref{eqn:q0stabcond}.
However, when we include first-order corrections (the attenuation terms $\mathfrak{a}_s$ and $\mathfrak{a}_{\mathfrak{D}}$), we exactly identify a similar complementary behavior between the sound mode and the diffusion mode.
We therefore find that the dynamical and thermodynamical analysis predict a similar behavior for the system, but point out that while the threshold value for the above thermodynamic quantities exactly corresponds to the point where the configuration obtains maximal charge density, the threshold \eqref{eqn:q0stabcond} for the dynamical stability corresponds to a smaller charge density as noted above (with the coincidentally exception of $n=1$).
A similar effect was also observed in \cite{Gath:2013qya}.
This can seem puzzling at first, but we emphasize that it is \emph{not} in contradiction with the correlated stability conjecture.




\subsection{A check: Mapping to AdS} \label{sec:map}

In the previous section we have analyzed the hydrodynamic behavior of dilatonic Maxwell charged brane solutions. In this section, we present a non-trivial check of our results with Ref.~\cite{Gouteraux:2011qh} using a modified version of the so-called ``AdS/Ricci-flat correspondence" \citep{Caldarelli:2012hy, *Caldarelli:2013aaa} which in its original form provides a map relating a certain class of asymptotically locally AdS solutions to Ricci-flat solutions. Starting with pure Einstein gravity, we modify the map by performing a Kaluza-Klein (KK) reduction. This gives us the possibility to connect with EMD theory, which is a subset of the class of theories \eqref{action}. 
From this starting point, we continue as Ref.~\citep{Caldarelli:2012hy, *Caldarelli:2013aaa} by performing a diagonal reduction and connect the reduced theory (after an appropriate analytic continuation) to theories which admit asymptotically local AdS solutions.
With this modification in hand, we can map the first-order corrected solutions of the previous section to their corresponding class of asymptotically locally AdS spacetimes obtained in Ref.~\cite{Gouteraux:2011qh}. 


We begin by performing a KK reduction of the Einstein-Hilbert action. This takes us to EMD theory and connects with the action \eqref{action} in $(n+p+3)$-dimensions with $q=0$ and fixes the coupling constant $a$ to
\begin{equation}\label{akk}
 a^2=\frac{2 (\alpha+1)}{\alpha} ~~,
\end{equation}
with $\alpha\equiv n+p+1$.
We then perform a diagonal reduction (over the sphere $S^{n+1}$)  with metric ansatz
\begin{equation}\label{flatansat}
\mathrm{d} s^2=e^{\frac{2}{\alpha}\chi(x,r)} \Big(\mathrm{d} s^2_{p+2}(x,r)+ \mathrm d \Omega^2_{(n+1)}\Big) ~~,
\end{equation}
where the $(p+2)$-dimensional reduced metric, the Maxwell field, the dilaton field, and the scalar field $\chi$ are independent of the $(n+1)$-directions of sphere. Note that this ansatz includes the solutions given by Eq.~\eqref{solutionSCHMaxwell}. The action of the lower dimensional theory takes the form
\begin{align}\label{Ricci4}
\mathcal{I}&=\mathrm{ Vol}(S^{n+1})\int_{p+2} e^{\chi} \bigg(\mathcal R\star 1+\star\thinspace n(n+1)-\frac{1}{\alpha(\alpha+1)}\mathrm d \phi\wedge \star\thinspace\mathrm d\phi\nonumber\\
&\qquad+\left(\frac{\alpha+1}{\alpha}\right)\mathrm d \chi\wedge\star\thinspace\mathrm d\chi
-\frac{1}{2}e^{-\frac{2}{\alpha}(\chi+\phi) } F_2\wedge\star\thinspace F_2\bigg) ~~,
\end{align}
where $\mathcal R$ is the $(p+2)$-dimensional Ricci scalar and $F_2$ is the field strength of the KK gauge field. Note that the second term comes from the integration over the sphere and that we have performed a suitable Weyl rescaling
\begin{equation}\label{rescal}
\phi^2 \rightarrow \frac{1}{2\alpha(\alpha+1)}\phi^2 ~~.
\end{equation}  


To establish the map, we now consider Einstein gravity with a negative cosmological constant $\Lambda=-d(d-1)/2$ in $(d+1)$-dimensions,\footnote{Working in units where the AdS radius is set to unity.}
\begin{equation}\label{AdS}
  I_{\Lambda}=\int_{d+1}\Big( R_{\Lambda}-2\Lambda\Big)*1 ~~.   
\end{equation}
We compactify the theory on a $\beta \equiv (d-p-1)$-dimensional torus using the following ansatz
\begin{equation}\label{AdSans}
\mathrm{d}s^2_{\Lambda}=\mathrm{d}s^2_{p+2}(x,r)+ e^{2\frac{\phi(x,r)+\chi(x,r)}{\beta} } \Big(\mathrm d y - A_a(x,r) \thinspace \mathrm d x^a \Big)^2+ e^{ \frac{2}{\beta}\left(\chi(x,r)-\frac{\phi(x,r)}{(\beta-1)}\right)}\mathrm d \ell^2_{\beta-1} ~~.
\end{equation}
Here $y$ is a distinguished direction along the torus $\mathbb T^{\beta}$ and $\mathrm d \ell^2_{\beta-1}$ denotes the line element of the remaining part of the torus.
Using the reduction ansatz \eqref{AdSans}, the reduced action is (dropping a total derivative)
\begin{align}\label{AdS1}
\mathcal{I}_{\Lambda}&=\mathrm{ Vol}(\mathbb T^\beta)\int_{p+2} e^{\chi}\bigg(\mathcal R \star 1+\star\thinspace d(d-1) -\frac{1}{\beta(\beta-1)} \mathrm d \phi\wedge\star\thinspace\mathrm d\phi\nonumber\\
&\qquad+\left(\frac{\beta-1}{\beta}\right)\mathrm d \chi\wedge\star\thinspace\mathrm d\chi
-\frac{1}{2}e^{2\frac{\phi+\chi}{\beta}}F_2\wedge\star\thinspace F_2\bigg) ~~,
\end{align}
with $F_2=\td A$.
Suppose that we now have access to closed analytic expressions of solutions to the reduced action \eqref{AdS1} for any positive integer $d$. It then makes sense to extend the domain via an analytic continuation \cite{Kanitscheider:2009as} of the parameter $d$. In this way, we can view solutions as a function of $d$ which now can take any real value. In particular, it makes sense to consider solutions for negative values of $d$. The same reasoning can be applied to the action \eqref{Ricci4} where the solutions are viewed as functions of $n$. 
Direct inspection of the two actions \eqref{AdS1} and \eqref{Ricci4} shows that they are proportional under the analytic continuation $d \leftrightarrow -n$,
\begin{equation} \label{eqn:actionmap}
	\mathrm{Vol}(S^{n+1}) \thinspace \mathcal{I}_\Lambda \; \leftrightarrow \; \mathrm{Vol(\mathbb{T^\beta})} \thinspace \mathcal{I} ~~.
\end{equation}
This implies that we can obtain solutions to the reduced theory \eqref{Ricci4} by a simple reflection of the dimensionality parameter starting from the reduced theory \eqref{AdS1} and vice versa.
In this way, knowing solutions of the form \eqref{flatansat} of the higher dimensional theory (for any $n \in \mathbb{N}$) allows us to consistently uplift to solutions of the form \eqref{AdSans} (for any $d-1 \in \mathbb{N}$) using the connection between the two reduced theories \eqref{eqn:actionmap} and vice versa.




Now, with the modified version of the mapping in hand, we can relate brane solutions of EMD theory + hydro to those of $\text{AdS}_{d+1}$ compactified on $(d-p-1)$ flat directions.\footnote{We note that care must be taken for fixing $n=1$ after the analytic continuation, since while this might still be some perturbed solution it will not correspond to the asymptotically flat perturbed solution that we have been considering. In a similar way one should take care from the reversed perspective if fixing $d=2$.}
The effective hydrodynamics of branes in $\text{AdS}_{d+1}$ compactified on a torus (as described above) was worked out in the Ref. \citep{Gouteraux:2011qh} directly from the AdS results in general $(d+1)$ dimensions \citep{Bhattacharyya:2008mz}.  We can thus map the solutions of the reduced AdS theory to solutions of EMD considered in this work and in turn compare the transport coefficients. This provides us with a non-trivial cross-check of our results for the particular value \eqref{akk} of the dilaton coupling. Going through this exercise, we find a perfect agreement between the transport coefficients (Eqs. \eqref{eqn:q0shearbulk}-\eqref{eqn:q0diffusion} and Eqs. (3.4.38)-(3.4.40) of \cite{Gouteraux:2011qh}, respectively.\footnote{Note, that \cite{Gouteraux:2011qh} uses a different normalization of the diffusion coefficient.}). 



Finally, we can in a similar way as in Sec.~\ref{sec:q0stability} check the dynamical features of the asymptotically local AdS solutions explicitly by computing the dispersion relations of the associated fluid. In contrast to the asymptotically flat solutions one finds that all the coefficients of both the sound mode \eqref{eqn:sosmode} and the diffusion mode \eqref{eqn:diffmode} are continuous and positive for all values of the charge density. We therefore find that no threshold value exists for the particular dilaton coupling \eqref{akk} and furthermore that the hyperbolic behavior has been resolved through the analytic continuation (compare e.g. with Fig.~\ref{fig:q0soundmode}). The solutions are thus stable to first order which is perhaps not too surprising, since they originate from stable neutral AdS branes \cite{Gregory:2000gf} through consistent reductions. In addition, one can easily check that they are also thermodynamically stable by computing the specific heat and isothermal permittivity (see Eq.~\eqref{eqn:q0specificheat}) and checking explicitly that they are positive for all values of the charge.

\section*{Discussion}
\addcontentsline{toc}{section}{Discussion}



We have performed the perturbative procedure which captures the hydrodynamic sector of fundamentally charged (dilatonic) black branes and branes with Maxwell charge smeared over their worldvolume.
The main result of this paper are the first-order transport coefficients that determine the dissipative behavior of the effective fluids. They are listed in Eq.~\eqref{transporttopform} and Eqs.~\eqref{eqn:q0shearbulk}-\eqref{eqn:q0diffusion}.
Furthermore, for each class of branes we have obtained the dispersion relations of the effective fluids for which we refer to Fig.~\ref{fig:qpsoundmode} and \ref{fig:q0soundmode}.
As explained in the main text, many of our results apply to brane descriptions arising in string and M-theory. We have highlighted the interesting values in Table~\ref{tab:g0critvales} and \ref{tab:q0table}.


We have computed the shear and bulk viscosities and in particular found their dependence on the (local) charge and dilaton coupling.
We find that both classes of black branes have a shear viscosity satisfying $\eta/s = 1/4\pi$ thus saturating the universality KSS bound \citep{Kovtun:2003wp, *Buchel:2003tz}. It is reasonable to expect that this result holds for any kind of smeared brane solution of the action \eqref{action}.
Even though there, at least to our knowledge, does not exist a universal bound on the bulk-to-shear viscosity ratio, it is nevertheless still interesting to test our result against the well-known (holographic) \quoting{bound} proposed by Buchel \cite{Buchel:2007mf}.
%
%
We find that the ratio for fundamentally charged branes always violates this bound for all non-zero values of the charge and only saturates it in the neutral limit. 
In contrast, we find that branes with smeared charge continue to satisfy the bound for small charge parameter, but break the bound for sufficiently large values of the charge parameter. 
Although violations of the Buchel bound already exist in the literature \cite{Buchel:2011uj}, our setup thus provides two additional and physically simple examples.
In Ref.~\cite{Gouteraux:2011qh} a different inequality and potential bound was proposed for  Maxwell charged AdS branes, but it is always violated for our asymptotically flat branes. This was also observed in \cite{DiDato:2013cla}.
Indeed, if a universal bound on the shear-to-bulk exists, it would need a rather a non-trivial modification, it seems.
%
%
Finally, the presence of smeared charge on the worldvolume gives rise to an additional hydrodynamic diffusion mode. We have determined the value of the transport coefficient $\mathfrak{D}$ associated with charge diffusion and explicitly found its dependence on the dilaton coupling. In fact, we note that even though it generalizes the result for zero dilaton coupling in \cite{Gath:2013qya}, it takes exactly the same form in terms of the parameter $N$.
We note that all the transport coefficients are positive as required by thermodynamic consistency (the second law of thermodynamics).


%

We have analyzed the dynamical stability of both classes of charged black branes by considering the response of their corresponding effective fluids to small long-wavelength perturbations. 
To leading order, both systems have a branch of configurations that suffers from a GL instability in the sound mode \citep{Gregory:1993vy, *Gregory:1994bj}. This branch connects naturally with the neutral brane configuration which is indeed known to be unstable \cite{Camps:2010br}. However, for a certain regime of the solution parameters, both systems also have a branch of stable configurations for a given charge connecting with the extremal limit.
When taking first-order effects into account, this seemingly similar behavior of the two systems breaks down.
%
%
%
For the system with fundamental charge we find that the attenuation coefficient of the sound mode is positive for all values of the charge and thus the stability is fully determined by the leading order i.e. by the speed of sound.
In contrast, an additional longitudinal diffusion mode plays a crucial r\^{o}le for the Maxwell system which in turn leads to a complementary behavior between the stability of the sound and diffusion mode.
The Maxwell black brane therefore suffers from a GL instability for all values of the charge.
This complementary behavior is similar to the behavior of thermodynamic stability under change of charge and hence our result is in accordance with the correlated stability conjecture \cite{Gubser:2000ec, Gubser:2000mm}.
We note that this was also observed for the Reissner-Nordstr\"om black brane considered in \cite{Gath:2013qya} and that similar observations have been done for other smeared branes, see e.g. \cite{Hirayama:2002hn, Ross:2005vh} for the case of smeared D$p$-branes.

\vspace{0.5cm}



In Sec.~\ref{sec:map}, we presented a cross-check of the transport coefficients associated with the fluid dynamics of the Maxwell charged branes utilizing a modified version of the \quoting{AdS/Ricci flat correspondence} \cite{Caldarelli:2012hy, Caldarelli:2013aaa}.
%
%
Although, it would be interesting to obtain a deeper understanding of whether any physics can be attributed to the existence of such a mapping, we can at least here comment with a partial answer with respect to its possible generalizations. 
In our endeavors trying to relate generic theories with an action of the type \eqref{action} to theories with a cosmological constant, we found that mappings of this kind are indeed possible. 
Unfortunately, despite allowing for a substantial amount of freedom between the field contents of the theories one in general ends up with pathological theories containing a scalar field having a kinetic term with the \quoting{wrong} sign. This is mainly due to severe restrictions arising from the requirements of consistent dimensional reductions.
However, it is worth mentioning an interesting application of the current map (as it is presented in Sec.~\ref{sec:map}), namely, that it provides us with a tool for working out the second-order corrected charged solutions by starting directly from the known second-order results in AdS \cite{Bhattacharyya:2008jc, Bhattacharyya:2008mz}. We leave this exercise for the future.
Before ending this paragraph, it is also interesting to note that since part of our computation is connected to the fluid description of the near horizon throat geometry (AdS) of stacks of D/M-branes, there are two, seemingly unrelated, schemes relating the results of the fluid/gravity correspondence to those of the asymptotically flat branes (and vice versa). It is therefore tempting to think that the two approaches are related although the precise connection still remains unclear. Hopefully our results will help shed some light over these unresolved issues.
In this regard it is also interesting to note that the hydrodynamic regimes of the spinning black D3-brane was recently considered in \cite{Erdmenger:2014jba} extending the work of \cite{Emparan:2013ila}.
Here, the theory naturally includes a $\text{U}(1)$ gauge field.
In the light of our results for the Maxwell black brane it would be of considerable interest to review whether they can be related (for specific parameters) to at least a subsector of the hydrodynamic limit of the spinning D3.

In this paper we have been concerned with the hydrodynamic sector of fundamentally charged black brane solutions and provided the dilaton generalization of the results obtained for the Reissner-Nordstr\"{o}m black brane \cite{Gath:2013qya}.
In order to put these results in a broader perspective it would be interesting to investigate the hydrodynamic limit for branes with general smeared charge ($q<p$) and possibly multi-charged bound state solutions. These results will be presented in \cite{genq}.
%
%
Furthermore, since smeared D0-brane solutions are related to charged D$p$-branes by T-duality, it would be interesting to work out the Buscher rules in a derivative expansion necessary to relate the derivative expansion of the solutions. Quite remarkably, one would then be able to utilize the map we have provided to obtain the second-order fluid transport for all D$p$-branes by simply starting from pure asymptotically local AdS solutions. We intent to pursue these ideas in the future.


\subsection*{Acknowledgments}
We would like to thank Roberto Emparan and Niels A. Obers for many useful discussions. In particular, Roberto Emparan for careful reading of the draft and his suggestions. ADD is also grateful to Jelle Hartong for many fruitful comments and discussions. We also thank the NBI for warm hospitality during various stages of the project. JG would also like to thank the hospitality of Barcelona University. ADD thanks the cooperation COST since part of this work was done during the COST Short Term Scientific Mission COST-STSM-MP1210-15746. JG is supported by VILLUM FONDEN, research grant VKR023371. AVP is supported by The Danish Council for Independent Research - Natural Sciences (FNU), DFF-4002-00307. ADD is supported by FPI scholarship BES-2011-045401 and grants MEC FPA2010-20807-C02-02, FPA2013-46570-C2-2-P, AGAUR 2009-SGR-168 and CPAN CSD2007-00042 ConsoliderIngenio 2010.

\appendix

\section{Details on the perturbative computation} \label{app:details}


For the convenience of the diligent reader, we here provide some of the details on our perturbative computation outlined in Sec. \ref{outlinepertexp}. 
When solving the perturbative equations we focus on the Maxwell charged case as it is in many regards the most intricate due to extra dynamical freedom in the vector sector. In subsection \ref{appsec:fund} we provide the most relevant equations (and differences) pertaining to the fundamentally charged case.

\subsection{Setting up the perturbative problem}

As explained in Sec. \ref{outlinepertexp}, in order for the pertubative problem to be well-posed, we need to cast the fields into (ingoing) Eddington-Finkelstein (EF) form $(x^a,r,\Omega)\to (\sigma^a,r,\Omega)$ with $\sigma^a=(v,\sigma^i)$, defined in the usual way, i.e., so that $|\td r|^2=0$. It is not difficult to verify that the coordinates, 
\begin{equation} \label{eqn:EFcoord}
\sigma^a=x^a+u^ar_\star \ , \quad r_\star(r)=r+\int_{r}^{\infty}\td r \thinspace \left(\frac{f-h^{N/2}}{f}\right) ~~,
\end{equation}
will do the job for both brane solutions \eqref{solutionSCH} and \eqref{solutionSCHMaxwell}. Note that we have chosen the integration constant appearing in $r_\star$ so that $r_\star\to r$ for large $r$. In this way the EF coordinates reduce to ordinary Schwarzschild light cone coordinates far from the horizon. Notice that it is possible to write down a closed form expression for $r_\star$ in terms of the hypergeometric Appell function $F_1$, 
\begin{equation}
r_\star(r)=r F_1 \left(-\frac{1}{n};-\frac{N}{2},1;1-\frac{1}{n};1-h,1-f\right)\approx r\left(1-\frac{1}{n-1} \frac{r_0^n}{r^n} \left(1+\frac{N\gamma_0}{2}\right)\right) ~~,
\end{equation}
where the last equality applies for large $r$ and is valid up to $\mathcal{O}\left(\frac{1}{r^{2n-1}}\right)$. With this definition of $r_\star$, we will limit our analysis to the case where $n\geq 2$ (the analysis for $n=1$ needs some modifications, however, at the end of the day, the results for the transport coefficients can be obtained by setting $n=1$ in the results derived below).

In EF coordinates \eqref{eqn:EFcoord}, the metric \eqref{solutionSCH} for the fundamentally charged brane takes the form 
\begin{equation}\label{eqn::qpbackground}
\td s^2=h^{-\frac{Nn}{p+n+1}}\Bigg(\Big( - f u_a u_b  +\Delta_{ab} \Big) \td \sigma^a \td \sigma^b - 2 h^{\frac{N}{2}}u_a \thinspace \td \sigma^a \td r + h^N r^2 \thinspace \td \Omega_{(n+1)}^2 \Bigg) \ .
\end{equation}
Similarly, the Maxwell charged brane \eqref{solutionSCHMaxwell} takes the form 
\begin{equation} \label{eqn::q0background}
\text{d}s^2=h^{-\left(\frac{n+p}{n+p+1}\right)N} \Bigg(\Big(  -f \thinspace u_a u_b + h^N \Delta_{ab} \Big) \td \sigma^a\td\sigma^b-2 h^{\frac{N}{2}} \thinspace u_a \thinspace \text{d}\sigma^a \text{d}r+h^{N}r^2\text{d}\Omega_{(n+1)}^2\Bigg)  ~~.
\end{equation}
Transforming to the coordinates \eqref{eqn:EFcoord} introduces non-zero radial components to the two gauge fields. This transformation can, however, be undone by a suitable gauge transformation and the EF form of the gauge fields and in this particular gauge read
\begin{equation}\label{gaugeEFcoord}
A = -\frac{1}{h}\left(\frac{r_0}{r}\right)^n\sqrt{N\gamma_0(\gamma_0+1)} \thinspace \star 1 \quad \quad \text{and} \quad \quad A=\frac{1}{h}\left(\frac{r_0}{r}\right)^n\sqrt{N\gamma_0(\gamma_0+1)} \thinspace u_a \text{d}\sigma^a ~~,  
\end{equation}
for the fundamentally and Maxwell charged branes, respectively. In particular, the two gauge fields only have components in the brane directions. Also note that the dilaton remains invariant under the coordinate transformation as it is independent of the brane directions.

We now promote the parameters $u^a$, $r_0$ and $\gamma_0$ to \emph{slowly} varying worldvolume fields and look for the corrections $\text{d}s^2_\partial$, $A_\partial$, $\phi_\partial$, so that $\td s^2_f$, $A_f$, $\phi_f$ solve the full set of EOMs to first order in the derivatives (cf. Eq. \eqref{fullsol}). In order to do this, we first need to expand the leading order seed solutions \eqref{eqn::qpbackground}-\eqref{gaugeEFcoord} to first order in the derivatives. Carrying out this expansion is of course straightforward, however, the resulting expressions are rather lengthy and not very illuminating and therefore we omit them here. Moreover, in the following we employ the inherent Lorentz symmetry of the background to work in the rest frame in the point $\mathcal P$ around which we consider hydrodynamic fluctuations (in the ultra-local sense). We therefore take $u^a(\sigma^a)|_{\mathcal P}=(1,\mathbf{0})$. Moreover, we define $r_0(\sigma^a)|_{\mathcal P}\equiv r_0$ and $\gamma_0(\sigma^a)|_{\mathcal P}\equiv \gamma_0$. Similarly, all the (worldvolume) derivatives are understood to be evaluated at $\mathcal P$. For example, $\partial_a r_0(\sigma^a)|_{\mathcal P}\equiv \partial_a r_0$ and so forth. 

The rest frame has a residual $\text{SO}(p)$ invariance which we use to split the resulting equations up into different sectors characterized according to their transformation under $\text{SO}(p)$. For the fundamentally charged black brane the scalar sector contains five scalars, the vector sector contains two vectors and the tensor sector contains one tensor. We parameterize the $\text{SO}(p)$ perturbations according to \vspace{2 mm} \begingroup
\setlength{\belowdisplayskip}{2pt} \setlength{\belowdisplayshortskip}{2pt}
\setlength{\abovedisplayskip}{2pt} \setlength{\abovedisplayshortskip}{2pt}
\begin{equation*} 
\begin{split}
\textbf{Scalar:} \ (A_{\partial})_{v i_1 ... i_p} &= -\sqrt{N\gamma_0(1+\gamma_0)} \thinspace \frac{r_0^n}{r^n} h^{-1} a_{v i_1 ... i_p} \ , \ \ (g_{\partial})_{vr}=h^{N\left(\frac{1}{2}-\frac{n}{n+p+1}\right)}f_{vr} \ , \\ 
(g_{\partial})_{vv} &= h^{1-\frac{N n}{n+p+1}}f_{vv} \ , \ \ \text{Tr} \thinspace (g_{\partial})_{ij}=h^{-\frac{N n}{n+p+1}} \text{Tr}f_{ij} \ , \ \ \phi_{\partial} = f_{\phi} \ ,
\end{split}
\end{equation*}
\begin{align} \label{eqn:pqdecom}
\textbf{Vector:} \ (g_{\partial})_{vi}=h^{-\frac{N n}{n+p+1}} f_{vi} \ , \ \ (g_{\partial})_{ri}=h^{N\left(\frac{1}{2}-\frac{n}{n+p+1}\right)}f_{ri} \ ,
\end{align}
\begin{equation*}
\textbf{Tensor:} \ (\overline{g}_{\partial})_{ij}=h^{-\frac{N n}{n+p+1}}\thinspace \overline{f}_{ij} \ , \vspace{2 mm}
\end{equation*} 
where $(\overline{g}_{\partial})_{ij}\equiv(g_{\partial})_{ij}-\frac{1}{p}(\text{Tr}(g_{\partial})_{kl}) \delta_{ij}$, i.e., the traceless part of $(g_{\partial})_{ij}$ and $ \overline{f}_{ij}\equiv f_{ij}-\frac{1}{p}(\text{Tr}f_{kl}) \delta_{ij}$. Similarly, the scalar sector of the Maxwell charged system contains five scalars while the vector and tensor sector contain three vectors and one tensor, respectively. Here we parameterize the $\text{SO}(p)$ perturbations in the three  sectors according to \vspace{2 mm}
\begin{equation*} 
\begin{split}
\textbf{Scalar:} \ & (A_{\partial})_v = -\sqrt{N\gamma_0(1+\gamma_0)} \thinspace \frac{r_0^n}{r^n} h^{-1} a_v \ , \ \ (g_{\partial})_{vr}=h^{N\left( - \frac{1}{2} + \frac{1}{n+p+1} \right)}f_{vr} \ , \\
 & (g_{\partial})_{vv}=h^{1-N\left(\frac{n+p}{n+p+1}\right)}f_{vv} \ , \ \ \text{Tr} \thinspace (g_{\partial})_{ij}=h^{\frac{N}{n+p+1}} \text{Tr}f_{ij} \ , \ \ \phi_{\partial} = f_{\phi} \ ,
\end{split}
\end{equation*}
\begin{align} \label{eqn:decom}
\textbf{Vector:} \ & (A_{\partial})_i = -\sqrt{ N \gamma_0(1+\gamma_0)} \thinspace a_i \ , \ \ (g_{\partial})_{vi}=h^{\frac{N}{n+p+1}} \thinspace f_{vi} \ , \\
 & \quad \quad \quad (g_{\partial})_{ri}=h^{N\left( - \frac{1}{2} + \frac{1}{n+p+1} \right)}f_{ri} \ , \nonumber
\end{align}
\begin{equation*}
\textbf{Tensor:} \ (\overline{g}_{\partial})_{ij}=h^{\frac{N}{n+p+1}} \thinspace \overline{f}_{ij} \ . \vspace{2 mm}
\end{equation*} 
As explained in the outline of Sec. \ref{outlinepertexp}, we will work in a gauge where all the rest of the components of $\td s^2_\partial$ and $A_\partial$ are taken to be zero. The consistency of this gauge choice can be checked a posteriori. \endgroup We now proceed with solving the resulting EOMs to first order in the derivatives. The EOMs (from the action \eqref{action}) take the form, 
\begin{equation}\label{EOMsapp}
\begin{split}
	R_{\mu\nu} - 2\nabla_{\mu}\phi\nabla_{\nu}\phi - S_{\mu\nu} &\equiv  \mathcal{E}_{\mu\nu} +\mathcal{O}(\partial^2) = 0 ~~, \\
	\nabla_{\mu} \left( e^{a\phi} \mathcal F^{\mu}_{\phantom{\mu} \rho_0 ... \rho_{q}} \right) &\equiv  \mathcal{M}_{\rho_0 ... \rho_{q}} +\mathcal{O}(\partial^2) = 0  ~~, \\
	 \quad g^{\mu\nu} \nabla_{\mu} \nabla_{\nu} \phi + \frac{a}{4(q+2)!} \mathcal F^2 &\equiv  \mathcal{E}_{(\phi)} +\mathcal{O}(\partial^2) = 0 ~~.
\end{split}
\end{equation}
Here $q=0$ for the Maxwell system and $q=p$ for the fundamentally charged system and $\mathcal F$ denotes the dilaton-weighted field strength $\mathcal F=e^{a\phi}\td A$. Moreover, we have defined
\begin{equation}
	S_{\mu\nu} = \frac{1}{2(q+1)!}  \left( \mathcal F_{\mu \rho_0 ... \rho_{q}} \mathcal F_{\nu}^{\phantom{\nu}\rho_0 ... \rho_{q}} - \frac{q+1}{(D-2)(q+2)} \mathcal F^2 g_{\mu\nu} \right) ~~.
\end{equation}
Notice that the right-hand sides of the EOMs \eqref{EOMsapp}, $\mathcal{E}_{\mu\nu}$, $\mathcal{M}_{\rho_0 ... \rho_{q}}$ and $\mathcal{E}_{(\phi)}$ are all $\mathcal{O}(\partial)$. The parameterization of the anz\"{a}tze \eqref{eqn:pqdecom} and \eqref{eqn:decom} are exactly chosen in such a way that the resulting equations only contain derivatives of $f_{ab}$, $a_{0\dotsc q}$ and $f_\phi$ and will thus be directly integrable as explained below Eq. \eqref{solgen}. In the two subsequent sections we provide the most important details for solving the two systems.

\subsection{Solving the Maxwell system}

In this section we give some of the details for solving the three $\text{SO}(p)$ sectors for the Maxwell charged system.

\subsubsection*{Scalars of \texorpdfstring{$\mathrm{SO}(p)$}{SO(p)}}


The scalar sector consists of eight independent equations which correspond to the vanishing of the components: $\mathcal{E}_{vv}, \mathcal{E}_{rv}, \mathcal{E}_{rr}, \text{Tr} \mathcal{E}_{ij}, \mathcal{E}_{\Omega\Omega}, \mathcal{E}_{(\phi)}, \mathcal{M}_v$ and $\mathcal{M}_r$ (cf. Eq. \eqref{EOMsapp}).



\paragraph{Constraint equations:}

There are two constraint equations; $\mathcal{E}^r_{\phantom{r}v}=0$ and $\mathcal{M}_r=0$. The two equations are solved consistently by
\begin{equation}
	\partial_v r_0 = - \frac{ r_0 ( 1 - (N-2)\gamma_0)}{ n+1 + (2-n(N-2)) \gamma_0 }  \partial_i u^i  \ , \quad \quad 
	\partial_v \gamma_0 = - \frac{2 \gamma_0(1+\gamma_0)} {n+1 + (2-n(N-2)) \gamma_0} \partial_i u^i  ~~.
\end{equation}
The first equation corresponds to conservation of energy while the second equation can be interpreted as current conservation. These are equivalent to the scalar conservation equations given by \eqref{eqn:conservation} in the rest frame. Under the assumption that the fluid configuration satisfies the above constraints one is left with six dynamical equations with five unknowns.

\paragraph{Dynamical equations:}

The coupled system constituted by the dynamical equations is quite intractable. One approach to obtaining the solution to the system is to decouple the trace function $\text{Tr} f_{ij}$. Once $\text{Tr} f_{ij}$ is known, it turns out, as will be presented below, that all the other functions can be obtained while ensuring that they are regular on the horizon.
It is possible to obtain a 3rd order ODE for $\text{Tr} f_{ij}$ by decoupling it through a number of steps. However, first it is useful to note that the particular combination of $\text{Tr}\mathcal{E}_{ij}$ and $\mathcal{E}_{(\phi)}$ leads to the equation
\begin{equation} \label{eqn:q0combieqn}
	\frac{\td}{\td r} \left[ r^{n+1} f(r) T'(r) \right] = - (\partial_i u^i) r^n  \left( 2(n+1) + C \frac{r_0^n}{r^n} \gamma_0 \right) h(r)^{\frac{N}{2} - 1} ~~,
\end{equation}
where we have defined the constant $C \equiv 2-n(N-2)$ and
\begin{equation} \label{eqn:trphi}
	T(r) = \text{Tr} f_{ij}(r) + \frac{4p}{(n+p+1) a} f_{\phi}(r) ~~.
\end{equation}
As we shall see this equation is very reminiscent of the equations for the tensor perturbations for which we know the solution to be
\begin{equation} \label{eqn:q0tsol}
	T(r) = c^{(1)}_T - 2(\partial_i u^i) \left( r_{\star} - \frac{r_0}{n} \left( 1 + \gamma_0 \right)^{\frac{N}{2}} \log f(r) \right) ~~.
\end{equation}
Here we have imposed horizon regularity, since $\text{Tr} f_{ij}$ and $f_{\phi}$ are individually regular on the horizon. Once $f_{\phi}$ is known in terms of $\text{Tr} f_{ij}$ we can use $\mathcal{E}_{rr}$ to eliminate $f'_{rv}$ and then take linear combinations of the remaining equations. The resulting combinations can then be used to eliminate $f'_{vv}$ and $f''_{vv}$ such that one is left with two equations in terms of $a_v$ and $\text{Tr} f_{ij}$ which can then be decoupled by standard means. The resulting equation is schematically of the form
\begin{equation}
	H^{(n,p)}_3(r) \left[ \text{Tr} f_{ij} \right] '''(r) + H^{(n,p)}_2(r) \left[ \text{Tr} f_{ij} \right]''(r) + H^{(n,p)}_1(r) \left[ \text{Tr} f_{ij} \right]'(r) = S_{\text{Tr}}(r) ~~,
	\label{eqn:dseqn}
\end{equation}
where $H_1, H_2$ and $H_3$ do not depend on the sources (world-volume derivatives) and the source term $S_{\text{Tr}}$ only depends on the scalar $\partial_i u^i$. The expressions for these functions are however very long and have therefore been omitted. After some work, one finds that the equation is solved by
\begin{equation} \label{eqn:tracesol}
	\text{Tr} f_{ij}(r) = c^{(1)}_{\text{Tr}} + \gamma_0 c^{(2)}_{\text{Tr}} G(r)  -  2(\partial_{i}u^{i}) \text{Tr} f_{ij}^{\text{(s)}}(r) ~~,
\end{equation}
where the terms containing the two integration constants $c^{(1)}_{\text{Tr}}$ and $c^{(2)}_{\text{Tr}}$ correspond to the homogeneous solution. The entire family of homogeneous solutions to equation \eqref{eqn:dseqn} of course has an additional one-parameter freedom which has been absorbed in the particular solution $\text{Tr} f_{ij}^{\text{(s)}}(r)$ and been used to ensure horizon regularity.\footnote{Note that equation \eqref{eqn:dseqn} has been derived under the assumption that $\partial_i u^i \neq 0$. This especially means that when there are no sources the one-parameter freedom disappears in accordance with \eqref{eqn:tracesol}.} With the introduction of $c^{(1)}_{\text{Tr}}$ we can safely take $c^{(1)}_{T}=0$. The function $G$ is given by
\begin{equation} \label{eqn:Goo1}
  G(r) = -\frac{N}{n+p+1} \frac{r_0^n}{r^n} \left(2 + \left(2 - \frac{Nn}{n+p+1}\right) \frac{r_0^n}{r^n} \gamma_0 \right)^{-1} ~~,
\end{equation}
and has an intricate relation to the gauge choice $(g_{\partial})_{\Omega\Omega} =0$ as we shall see later. The particular solution which is regular on the horizon is given by
\begin{equation} \label{eqn:tracesol2}
	\text{Tr} f_{ij}^{\text{(s)}}(r) = \frac{r_0}{n}  (1+\gamma_0)^{\frac{N}{2}} \alpha \gamma_0 G(r) + \left( r_{\star} - \frac{r_0}{n} \left( 1 + \gamma_0 \right)^{\frac{N}{2}} \log f(r) \right) \left( 1 + \beta \gamma_0  G(r) \right) ~~,
\end{equation}
with the coefficients
\begin{equation}
	\alpha = 2 p \left( \frac{ 2 (n+1) +  C \gamma_0}{(n+1)^2 + C \gamma_0 (2 (n+1) + C \gamma_0)} \right) \quad \text{and} \quad
	\beta = p \left( \frac{n+2 + C\gamma_0}{n+1 + C\gamma_0} \right)  ~~.
\end{equation}
With $\text{Tr} f_{ij}$ (and $f_{\phi}$) given, the equation $\mathcal{E}_{rr}=0$ will provide the derivative of $f_{rv}$,
\begin{equation} 
	f'_{rv}(r) = \frac{r}{\left(2(n+1) + C \frac{r_0^n}{r^n} \gamma_0 \right) h(r)^{\frac{N}{2}-1}} \left( \frac{\td}{\td r}\left[ h(r)^{\frac{N}{2}} [\text{Tr} f_{ij}]'(r) \right]  - 4a\frac{\td}{\td r}\left[ h(r)^{\frac{N}{2}} \right] f'_{\phi}(r) \right)  ~~. \label{eqn:dfrv}
\end{equation}
Since this equation is a 1st order ODE, the regularity of the horizon is ensured by $\text{Tr} f_{ij}$. Note that it is possible to perform integration by parts and use that the derivative of $r_{\star}$ takes a simpler form. One can thereafter obtain an analytical expression for the resulting integral. This expression is, however, quite cumbersome and we therefore only provide its large $r$ asymptotics
\begin{equation}
	f_{rv}(r) \approx f^{\text{(h)}}_{rv}(r) +
	(\partial_i u^i) \sum_{k=1}^{\infty} \frac{r_0^{nk}}{r^{nk}} \left[  \alpha^{(k)}_{rv} r + \beta^{(k)}_{rv} r_0  \right]  ~~.
\end{equation}
Here the homogeneous solution takes the form
\begin{equation}
	f^{\text{(h)}}_{rv}(r) = c_{rv} +
	\gamma_0 N c^{(2)}_{\text{Tr}}  \frac{r_0^n}{r^n} \left(  \frac{ 2p(n+p+1) + (n+p)(2p+C)\frac{r_0^n}{r^n} \gamma_0 }{2p \left( 2(n+p+1) + (2p+C) \frac{r_0^n}{r^n} \gamma_0 \right)^2} \right) ~~,
\end{equation}
and the particular solution is given in terms of the coefficients $\alpha^{(k)}_{rv}$ and $\beta^{(k)}_{rv}$ which depend on $n,p,a$, and $\gamma_0$. These coefficients are in general very long and their expressions are therefore omitted.

Using the expression for $f'_{rv}$ in terms of $\text{Tr} f_{ij}$, the Maxwell equation $\mathcal{M}_v=0$ becomes a 2nd order ODE for the gauge field perturbation,
\begin{equation} 
	\frac{\td}{\td r}\left[ \frac{1}{r^{n-1}} a'_v(r) \right] = \frac{n r^2}{ \left(2(n+1) + C \frac{r_0^n}{r^n} \gamma_0 \right)} \frac{\td}{\td r}\left[ \frac{1}{r^{n+1}} [\text{Tr} f_{ij}]'(r) + 4a\frac{(n+1)}{r^{n+2}}f_{\phi}'(r) \right] ~~. \label{eqn::av}
\end{equation}
This equation is solved by a double integration. The inner integral is manifestly regular at the horizon, one can therefore work directly with the asymptotic behavior of the right-hand side before performing the integrations. The large $r$ behavior of the perturbation function is thus found to be
\begin{equation}
	a_{v}(r) \approx a^{\text{(h)}}_{v}(r) + (\partial_i u^i) \left( - \frac{n}{n-1} r + \sum_{k=1}^{\infty} \frac{r_0^{nk}}{r^{nk}} \left[ \alpha_v^{(k)} r + \beta_v^{(k)} r_0 \right] \right) ~~,
\end{equation}
where the first term constitute the homogeneous solution,
\begin{equation}
	a^{\text{(h)}}_{v}(r) = c_v^{(1)} r^n + c_v^{(2)} - \gamma_0 c^{(2)}_{\text{Tr}}  \frac{r_0^n}{r^n} \left(  \frac{ 2p+C }{2p \left( 2(n+p+1) + (2p+C) \frac{r_0^n}{r^n} \gamma_0 \right)} \right)  ~~,
\end{equation}
and the particular solution is given in terms of the coefficients $\alpha^{(k)}_{v}$ and $\beta^{(k)}_{v}$ depending on $n,p,a$, and $\gamma_0$.

The last perturbation function $f_{vv}$ can be obtained from $\text{Tr}\mathcal{E}_{ij}=0$ which provides a 1st order ODE for the perturbation.  Horizon regularity is therefore ensured by the horizon regularity of $\text{Tr} f_{ij}$. Using the expression for $f'_{rv}$ in terms of $\text{Tr} f_{ij}$ the equation is schematically of the form
\begin{equation}
	f'_{vv}(r) = G_1\left[ \text{Tr} f_{ij}(r) \right] + G_2\left[a_v(r) \right] + G_3\left[f_{\phi}(r) \right]  + S_{ii}(r)  ~~, \label{eqn:fvv}	
\end{equation}
where $G_1, G_2, G_3$ are differential operators and the source $S_{ii}$ depends on $\partial_i u^i$. Again, the full expressions have been omitted and we only provide the large $r$ behavior,
\begin{equation}
	f_{vv}(r) \approx f^{\text{(h)}}_{vv}(r) + (\partial_i u^i)
	\sum_{k=1}^{\infty} \frac{r_0^{nk}}{r^{nk}} \left[ \alpha_{vv}^{(k)} r + \beta_{vv}^{(k)} r_0 \right]	~~,
\end{equation}
with the homogeneous part given by
\begin{equation}
\begin{split}
	f^{\text{(h)}}_{vv}(r) =& c_{vv}^{(1)} + \frac{r_0^{n}}{r^{n}} \frac{1}{h(r)} \Bigg( -2(1+\gamma_0) (c_v^{(2)} - c_v^{(1)}r_0^n \gamma_0) 
	+ c^{(1)}_{\text{Tr}} \frac{(n+p+1) (1+\gamma_0) a^2}{2p} \\	
	& + \Bigg(  \frac{ (n+p+1 +(2p+C)\gamma_0)h(r) - p\gamma_0 Nf(r) }{p \left( 2(n+p+1) + (2p+C) \frac{r_0^n}{r^n} \gamma_0 \right)} \Bigg) c^{(2)}_{\text{Tr}} \Bigg) ~~,
\end{split}
\end{equation}
and where the coefficients $\alpha^{(k)}_{vv}$ and $\beta^{(k)}_{vv}$ again depend on $n,p,a$, and $\gamma_0$.

Finally, one must ensure that the remaining equations coming from $\mathcal{E}_{vv}$ and the angular directions ($\mathcal{E}_{\Omega\Omega}=0$) are satisfied. This will require the following relation
\begin{equation}
	c^{(1)}_{vv} = - 2 c_{rv} ~~. \\
\end{equation}
This completes the analysis of the scalar sector. The remaining undetermined integration constants are thus: $c^{(1)}_{\text{Tr}}, c^{(2)}_{\text{Tr}}, c_{rv}, c_v^{(1)}, c_v^{(2)}$. Note that the above functions reproduce the neutral case as $\gamma_0 \rightarrow 0$.

\subsubsection*{Vectors of \texorpdfstring{$\mathrm{SO}(p)$}{SO(p)}}

The vector sector consists of $3p$ independent equations which correspond to the vanishing of the components: $\mathcal{E}_{ri}, \mathcal{E}_{vi}$ and $\mathcal{M}_i$.

\paragraph{Constraint equations:}

The constraint equations are given by the Einstein equations $\mathcal{E}^r_{\phantom{r} i}=0$ and are solved by
\begin{equation} \label{eqn:consVecSec}
	\partial_i r_0 = r_0 (1 + N \gamma_0) \partial_v u_i ~~,
\end{equation}
which are equivalent to conservation of stress-momentum. These are part of the conservation equations given by \eqref{eqn:conservation} in the rest frame. Similar to above we now proceed solving for the first-order corrections to the metric and gauge field under the assumption that the fluid profile satisfies the above constraint \eqref{eqn:consVecSec}.

\paragraph{Dynamical equations:}

The remaining equations consist of $p$ pairs of one Einstein equation $\mathcal{E}_{v i}=0$ and one Maxwell equation $\mathcal{M}_{i}=0$. The structure of these equations is the same as in the scalar sector. The Einstein equation $\mathcal{E}_{vi}=0$ is schematically of the form,
\begin{equation}
	L_3^{(n,p)}(r) f_{vi}''(r) + L_2^{(n,p)}(r) f_{vi}'(r) + L_1^{(n,p)}(r) a_i'(r)  = S_{vi}(r) ~~,
\end{equation}
while the Maxwell equation $\mathcal{M}_{i}=0$ is,
\begin{equation}
	M_3^{(n,p)}(r) a_i''(r) + M_2^{(n,p)}(r) a_i'(r) + M_1^{(n,p)}(r) f_{vi}'(r) = S_{i}(r) ~~.
\end{equation}
We omit the expressions for the functions $L_k$ and $M_k$, $k=1,...,3$.

To decouple the system we differentiate $\mathcal{E}_{vi}$ once and eliminate all $a_i(r)$ terms in $\mathcal{M}_{i}$. Doing so, one obtains a 3rd order ODE for $f_{vi}(r)$ which can be written on the form
\begin{equation}
	\frac{\td }{\td r} \left[ \frac{r^{n+1} f(r)}{h^{N}} \left( 1 - c_1 \frac{r_0^n}{r^n} \right)^2 \frac{\td}{\td r} \left[ \frac{ r^{n+1} h^{N+1}}{ \left( 1 - c_1 \frac{r_0^n}{r^n} \right) } f'_{vi}(r) \right] \right] 
	= S_{vi}(r) \ ,  \ \ \quad c_1 \equiv \frac{N-1}{1+ N\gamma_0}\gamma_0 \ .
\end{equation}
It is possible to perform the first two integrations analytically and ensure regularity at the horizon. The first integration is straightforward while the second involves several non-trivial functions. The large $r$ behavior of the $f_{vi}$ function is found to be
\begin{equation} \label{eqn:q0fvi}
	f_{vi}(r) \approx c^{(1)}_{vi} - \left( 1- \frac{f(r)}{h(r)^N} \right) c_{vi}^{(2)} - (\partial_v u_i) r + \sum_{k=1}^{\infty} \frac{r_0^{nk}}{r^{nk}} \left[ \alpha^{(k)}_{vi} r  + \beta^{(k)}_{vi} r_0 \right]  ~~,
\end{equation}
where the first two terms constitute the homogeneous solution and we find, in particular, that in order to ensure horizon regularity one must have
\begin{equation}
	\beta^{(2)}_{vi} = - \frac{N}{4n} \left( \frac{2\gamma_0(1+\gamma_0) (\partial_v u_i) + (\partial_i \gamma_0)}{(1+\gamma_0)^{\frac{N}{2}-1} (1 + N \gamma_0) } \right) ~~.
\end{equation}
The remaining set of coefficients $\alpha^{(k)}_{vi}$ and $\beta^{(k)}_{vi}$ are in general complicated expressions depending on the parameters in the problem. We therefore omit them as they provide no insight. Also, we notice that the sum in the function \eqref{eqn:q0fvi} vanishes in the neutral limit.

Once the solution of $f_{vi}$ is given we can use $\mathcal{E}_{vi}$ to determine $a_i$,
\begin{equation}
	a_{i}(r) \approx  c^{(1)}_{i} + \frac{r_0^n}{r^n}\frac{1}{h(r)} c_{vi}^{(2)} + \sum_{k=1}^{\infty} \frac{r_0^{nk}}{r^{nk}} \left[ \alpha^{(k)}_{i} r  + \beta^{(k)}_{i} r_0 \right] ~~,
\end{equation}
where the first two terms correspond to the homogeneous solution. Again, we choose to omit the coefficients $\alpha^{(k)}_{i}$ and $\beta^{(k)}_{i}$. The remaining undetermined integration constants are thus: $c^{(1)}_{i}, c^{(1)}_{vi}$, and $c^{(2)}_{vi}$.

\subsubsection*{Tensors of \texorpdfstring{$\mathrm{SO}(p)$}{SO(p)}} \label{sec:q0tensorsector}

There are no constraint equations in the tensor sector which consists of $p(p+1)/2-1$ dynamical equations given by
\begin{equation}
	\mathcal{E}_{ij} - \frac{\delta_{ij}}{p} \text{Tr}(\mathcal{E}_{ij}) = 0 ~~.
\end{equation}
This gives an equation for each component of the traceless symmetric perturbation functions $\bar{f}_{ij}$,
\begin{equation} \label{eqn:q0:tensoreqn}
	\frac{\td}{\td r} \left[ r^{n+1} f(r) \bar{f}'_{ij}(r) \right] = - \sigma_{ij} r^n  \left( 2(n+1) + C \frac{r_0^n}{r^n} \gamma_0 \right) h(r)^{\frac{N}{2} - 1} ~~,
\end{equation}
with the same structure as Eq.~\ref{eqn:q0combieqn}. The spatial part of the shear tensor is
\begin{equation} \label{eqn:spatialsheartensor}
	\sigma_{ij} = \partial_{(i}u_{j)} - \frac{1}{p}\delta_{ij} \partial_k u^k ~~.
\end{equation}
The solution is given by,
\begin{equation}
	\bar{f}_{ij}(r) = \bar{c}_{ij} - 2\sigma_{ij} \left( r_{\star} - \frac{r_0}{n} \left( 1 + \gamma_0 \right)^{\frac{N}{2}} \log f(r) \right) ~~,
\end{equation}
where horizon regularity has been imposed and the integration constant(s) $\bar{c}_{ij}$ is symmetric and traceless.
\subsection{Solving the fundametally charged system}\label{appsec:fund}
As already explained, many of the differential equations appearing in fundamentally charged system are similar to the ones appearing for the Maxwell system. Instead of repeating these, in this section we provide the most important differences.

\subsubsection*{Scalars of \texorpdfstring{$\mathrm{SO}(p)$}{SO(p)}}

The scalar sector consists of eight independent equations which correspond to the vanishing of the components: $\mathcal{E}_{vv}, \mathcal{E}_{rv}, \mathcal{E}_{rr}, \text{Tr} \mathcal{E}_{ij}, \mathcal{E}_{\Omega\Omega}, \mathcal{E}_{(\phi)}, \mathcal{M}_{a_1...a_{p+1}}$ and $\mathcal{M}_{r i_1 ... i_p}$ with $i \in \{ \sigma^i \}$.

\paragraph{Constraint equations:}

There are two constraint equations; $\mathcal{E}^r_{\phantom{r}v}=0$ and $\mathcal{M}_{r i_1 \ldots i_p}=0$ with $i \in \{ \sigma^i \}$. The two equations are solved consistently by
\begin{equation} \label{eqn:pqconsScalarSec1}
	\partial_v r_0 = - \frac{ r_0 ( 1 + 2\gamma_0 )}{ n+1 + (2-n(N-2)) \gamma_0 }  \partial_i u^i  \ , \quad \quad	\partial_v \gamma_0 = - \frac{2 n \gamma_0(1+\gamma_0)} {n+1 + (2-n(N-2)) \gamma_0} \partial_i u^i  ~~.
\end{equation}
The first equation corresponds to conservation of energy while the second equation can be interpreted as the charge density being constant in time.

\paragraph{Dynamical equations:}

After the constraint Eqs.~\eqref{eqn:pqconsScalarSec1} have been imposed one is left with a system very similar to the one obtained in the presence of Maxwell charge ($q=0$). It consists of the six equations given by the components: $\mathcal{E}_{vv}$, $\mathcal{E}_{rr}$, $\text{Tr}\mathcal{E}_{ij}$, $\mathcal{E}_{\Omega\Omega}$, $\mathcal{E}_{(\phi)}$, and $\mathcal{M}_{a_1 ... a_{p+1}}$.

The particular combination of $\text{Tr}\mathcal{E}_{ij}$ and $\mathcal{E}_{(\phi)}$ gives Eq.~\eqref{eqn:q0combieqn}, but now with the relation
\begin{equation} \label{eqn:pqtphi}
	T(r) = \text{Tr} f_{ij}(r) - \frac{4np}{(n+p+1)a} f_{\phi}(r) ~~,
\end{equation}
and is solved by the same expression given by Eq.~\eqref{eqn:q0tsol}. The equation for the trace $\text{Tr} f_{ij}$ is again similar to the $q=0$ case and the solution can be put on the form given by \eqref{eqn:tracesol} with \eqref{eqn:tracesol2}, but where
\begin{equation} \label{eqn:Goo2}
  G(r) = -\frac{N(p+1)}{n+p+1} \frac{r_0^n}{r^n} \left(2 + \left(2 - \frac{Nn(p+1)}{n+p+1}\right) \frac{r_0^n}{r^n} \gamma_0 \right)^{-1} ~~,
\end{equation}
and the coefficients are given by the expressions
\begin{equation}
	\alpha = \frac{2p n^2}{p+1} \left( \frac{ 2 (n+1) +  C \gamma_0}{(n+1)^2 + C \gamma_0 (2 (n+1) + C \gamma_0)} \right) \quad \text{and} \quad
	\beta = \frac{p}{p+1} \left( \frac{n^2}{n+1 + C\gamma_0} \right)  ~~.
\end{equation}
The solution of $\text{Tr}f_{ij}$ dictates the perturbation of the dilaton field through Eq.~\eqref{eqn:pqtphi}. With $\text{Tr}f_{ij}$ determined, one can find the remaining perturbation functions as follows: $f_{rv}$ from $\mathcal{E}_{rr}$ and $f_{vv}$ from $\text{Tr} \mathcal{E}_{ij}$ by a single integration while the gauge field perturbation $a_{v i_1 ... i_p}$ can be obtained from $\mathcal{M}_{v a_1 ... a_p}$ by a double integration.
Finally, we note that remaining undetermined integration constants are equivalent to the $q=0$ case.

\subsubsection*{Vectors of \texorpdfstring{$\mathrm{SO}(p)$}{SO(p)}}

The vector sector consists of $3p$ independent equations which correspond to the vanishing of the components: $\mathcal{E}_{ri}$, $\mathcal{E}_{vi}$ and $\mathcal{M}_{vr j_1 \ldots j_{p-1}}$ with $ j \neq i$.

\paragraph{Constraint equations:}

The constraint equations are given by the Einstein equations $\mathcal{E}^r_{\phantom{r} i}=0$ and $\mathcal{M}_{vr j_1 \ldots j_{p-1}}=0$ with $ j \neq i$. For each spatial index $i$ one has a pair of equations that are solved by
\begin{equation} \label{eqn:pqconsVecSec1}
	\partial_i r_0 = \frac{r_0 (1+2\gamma_0) }{ 1 - (nN-2) \gamma_0 } \partial_v u_i \ , \quad \quad 	\partial_i \gamma_0 = - \frac{2n \gamma_0 (1+\gamma_0) }{ 1 - (nN-2)\gamma_0 } \partial_v u_i ~~.
\end{equation}
The first equation corresponds to conservation of stress-momentum while the second equation censures that the charge density does not have any spatial gradients over the world-volume. We see that the current is more constrained compared to the case with Maxwell charge which is tied to the fact that the $p$-brane charge is not able to redistribute itself.

\paragraph{Dynamical Equations:}
 After the constraint Eqs.~\eqref{eqn:pqconsVecSec1} have been imposed the remaining $p$ equations consist of 2nd order differential equations $\mathcal{E}_{v i}=0$ of the form
\begin{equation}	
	\frac{\td}{\td r} \left[ r^{n+1} f'_{vi}(r) \right] = S_{vi}(r) ~~.
\end{equation}
Each equation can be integrated analytically in order to obtain the perturbation functions $f_{vi}$. Horizon regularity is ensured due to the form of the differential operator. We note that the homogeneous solution gives rise to two integration constants: $c^{(1)}_{vi}$ and $c^{(2)}_{vi}$.

\subsubsection*{Tensors of \texorpdfstring{$\mathrm{SO}(p)$}{SO(p)}}

It turns out that with the parametrization given by Eq.~\eqref{eqn:pqdecom}, the equations for the tensor perturbations take the exact same form as found for the Maxwell charge given by the form \eqref{eqn:q0:tensoreqn}. The solution is therefore,
\begin{equation}
	\bar{f}_{ij}(r) = \bar{c}_{ij} - 2\sigma_{ij} \left( r_{\star} - \frac{r_0}{n} \left( 1 + \gamma_0 \right)^{\frac{N}{2}} \log f(r) \right) ~~,
\end{equation}
with $\sigma_{ij}$ given by Eq.~\eqref{eqn:spatialsheartensor} and regularity of the horizon has been imposed. The constant $\bar{c}_{ij}$ is again symmetric and traceless. Because of this closed-form expression, the shear viscosity will take the same form as found for the system with Maxwell charge.

\subsection{Fixing the integration constants}
We have determined the first-order derivative corrected solution for both types of branes. Both solutions are completely determined up to a set of integration
constants, $c_{\text{Tr}}^{(1)}$, $c_{\text{Tr}}^{(2)}$, $c_v^{(1)}$, $c_v^{(2)}$, $c_{vi}^{(1)}$, $c_{vi}^{(2)}$, $c_{rv}$, $\bar c_{ij}$ (and $c_i^{(1)}$ for $q=0$). As explained in Sec. \ref{outlinepertexp}, these constants are fixed by virtue of asymptotic flatness and choice of fluid frame (gauge). The latter freedom exactly corresponds to constant $\mathcal{O}(\partial)$ shifts of the parameters of the $\mathcal{O}(\partial^0)$ fields, i.e., it parameterizes the homogeneous solution to the above differential equations. Indeed, for the zeroth order Maxwell solution \eqref{eqn::q0background}, consider order $\mathcal{O}(\partial)$ constant shifts $r_0 \to r_0 + \delta r_0$, $\gamma_0 \to \gamma_0 +  \delta \gamma_0$ and a constant gauge shift $a_v \to a_v + \delta a_v$. Moreover, by redefining the $r$ coordinate as,
\begin{equation} \label{eqn:rcoordshift}
	r \rightarrow r \left( 1 + \gamma_0 (n \delta \log r_0 + \delta \log \gamma_0) G(r) \right) ~~,
\end{equation}
with $G(r)$ given by \eqref{eqn:Goo1}, the angular directions do not receive first-order contributions in accordance with the gauge choice $(g_{\partial})_{\Omega\Omega}=0$. The resulting expressions after the shifts and coordinate transformation are of course still solutions and the overall change exactly corresponds to the homogeneous solution to the above differential equations in the scalar sector. More precisely, one can relate the integration constants to the shifts by,
\begin{equation}
\begin{split} \label{eqn:relations}
	\ \ &c^{(2)}_{\text{Tr}} = - 2p(n \delta \log r_0 + \delta \log \gamma_0) \ , \quad c^{(1)}_v = -\frac{\delta a_v}{r_0^n \sqrt{N \gamma_0 (1+\gamma_0)}} \ , \\
&\ \ c^{(2)}_v = -n\delta \log r_0  - \frac{1+2\gamma_0}{2(1+\gamma_0)} \delta \log \gamma_0 - \frac{\gamma_0}{\sqrt{N \gamma_0 (1+\gamma_0)}} \delta a_v  \ .
\end{split} 
\end{equation}For the vector sector one finds that the homogeneous part of the above solution corresponds to global constant shift of the boost $u_i \rightarrow u_i + \delta u_i$ and  in the gauge $a_i \rightarrow a_i + \delta a_i$. With the same choice of radial coordinate, one has
\begin{equation}
	c^{(2)}_{vi} =  \delta u_i \ , \quad 
	c^{(1)}_{i} = -\frac{\delta a_i}{\sqrt{N \gamma_0 (1+\gamma_0)}} \ .
\end{equation}
Similar expressions relating the shifts to the integration constants can be derived for the fundamentally charged solution. In the following we require all the $\mathcal{O}(\partial)$ shifts to vanish. In the effective fluid description this exactly corresponds to choosing the Landau frame. This fixes the integration constants, $c^{(2)}_{\text{Tr}}=c^{(1)}_v= c^{(2)}_v=c^{(2)}_{vi} =0$ (and $c^{(1)}_{i}=0$ for $q=0$).

To fix the remaining integration constants, we now impose asymptotic flatness to $\mathcal O (\partial)$. In order to do so, we must first transform our results back into Schwarzschild form. In addition to making the asymptotics more transparent, this is also needed for extracting the effective hydrodynamic currents. In order to change coordinates, we use the inverse of the transformation stated in equation \eqref{eqn:EFcoord} to $\mathcal O (\partial)$. The inverse transformation is worked out iteratively order by order. To first order, the transformation from EF to Schwarzschild coordinates is found to be,
\begin{align} \label{eqn:invcoordtrans}
	v &= t + r_{\star} +  \bigg[ (t + r_{\star}) \left( \partial_{r_0} r_{\star} \partial_t r_0 + \partial_{\gamma_0} r_{\star} \partial_t \gamma_0  \right) +  x^i \left( \partial_{r_0} r_{\star} \partial_i r_0 + \partial_{\gamma_0} r_{\star} \partial_i \gamma_0  \right) \bigg] + \mathcal{O}(\partial^2) ~~, \nonumber\\
	\sigma^i &= x^i +  \bigg[ (t+r_{\star}) \partial_t u^i + \sigma^j \partial_j u^i \bigg] r_{\star} + \mathcal{O}(\partial^2) ~~.
\end{align}
It is now possible to express all the fields in Schwarzschild coordinates and impose asymptotic flatness. This leads to \begin{equation}
c_{rv} = 0 \ , \quad c^{(1)}_{vi}=0  \ , \quad c^{(1)}_{\text{Tr}} = 0  \ , \quad \bar{c}_{ij} = 0 \ , \end{equation} 
for both types of branes. Having obtained the full first-order derivative corrected asymptotically flat solutions for both the Maxwell and fundamentally charged brane, it is now possible to read off the effective hydrodynamic currents using standard methods (see also \cite{Gath:2013qya}). Our results for the transport coefficients are presented in Sec.~\ref{sec:qptransport} and \ref{sec:q0transport}.

\newpage

\addcontentsline{toc}{section}{References}
\bibliographystyle{newutphys}
\bibliography{References}

\providecommand{\href}[2]{#2}\begingroup\raggedright\begin{thebibliography}{10}

\bibitem{Polchinski:1995mt}
J.~Polchinski, ``{Dirichlet Branes and Ramond-Ramond charges},''
  \href{http://dx.doi.org/10.1103/PhysRevLett.75.4724}{{\em Phys.Rev.Lett.}
  {\bf 75} (1995)  4724--4727},
\href{http://arxiv.org/abs/hep-th/9510017}{{\tt arXiv:hep-th/9510017
  [hep-th]}}.

\bibitem{Maldacena:1997re}
J.~M. Maldacena, ``{The Large N limit of superconformal field theories and
  supergravity},'' \href{http://dx.doi.org/10.1023/A:1026654312961}{{\em
  Int.J.Theor.Phys.} {\bf 38} (1999)  1113--1133},
\href{http://arxiv.org/abs/hep-th/9711200}{{\tt arXiv:hep-th/9711200
  [hep-th]}}.

\bibitem{Damour:1978cg}
T.~Damour, ``{Black Hole Eddy Currents},''
\href{http://dx.doi.org/10.1103/PhysRevD.18.3598}{{\em Phys.Rev.} {\bf D18}
  (1978)  3598--3604}.

\bibitem{Price:1986yy}
R.~Price and K.~Thorne, ``{Membrane Viewpoint on Black Holes: Properties and
  Evolution of the Stretched Horizon},''
\href{http://dx.doi.org/10.1103/PhysRevD.33.915}{{\em Phys.Rev.} {\bf D33}
  (1986)  915--941}.

\bibitem{Bhattacharyya:2008jc}
S.~Bhattacharyya, V.~E. Hubeny, S.~Minwalla, and M.~Rangamani, ``{Nonlinear
  Fluid Dynamics from Gravity},''
  \href{http://dx.doi.org/10.1088/1126-6708/2008/02/045}{{\em JHEP} {\bf 0802}
  (2008)  045},
\href{http://arxiv.org/abs/0712.2456}{{\tt arXiv:0712.2456 [hep-th]}}.

\bibitem{landau_fluid_1987}
L.~D. Landau and E.~M. Lifshitz, {\em Fluid mechanics}.
\newblock Course of Theoretical Physics, Vol. 6. Pergamon Press, London,
  2nd~ed., 1987.
\newblock Translated from the Russian by J. B. Sykes and W. H. Reid.

\bibitem{Policastro:2001yc}
G.~Policastro, D.~Son, and A.~Starinets, ``{The Shear viscosity of strongly
  coupled N=4 supersymmetric Yang-Mills plasma},''
  \href{http://dx.doi.org/10.1103/PhysRevLett.87.081601}{{\em Phys.Rev.Lett.}
  {\bf 87} (2001)  081601},
\href{http://arxiv.org/abs/hep-th/0104066}{{\tt arXiv:hep-th/0104066
  [hep-th]}}.

\bibitem{Maeda:2006by}
K.~Maeda, M.~Natsuume, and T.~Okamura, ``{Viscosity of gauge theory plasma with
  a chemical potential from AdS/CFT},''
  \href{http://dx.doi.org/10.1103/PhysRevD.73.066013}{{\em Phys.Rev.} {\bf D73}
  (2006)  066013},
\href{http://arxiv.org/abs/hep-th/0602010}{{\tt arXiv:hep-th/0602010
  [hep-th]}}.

\bibitem{Son:2006em}
D.~T. Son and A.~O. Starinets, ``{Hydrodynamics of r-charged black holes},''
  \href{http://dx.doi.org/10.1088/1126-6708/2006/03/052}{{\em JHEP} {\bf 0603}
  (2006)  052},
\href{http://arxiv.org/abs/hep-th/0601157}{{\tt arXiv:hep-th/0601157
  [hep-th]}}.

\bibitem{Banerjee:2008th}
N.~Banerjee, J.~Bhattacharya, S.~Bhattacharyya, S.~Dutta, R.~Loganayagam, {\em
  et al.}, ``{Hydrodynamics from charged black branes},''
  \href{http://dx.doi.org/10.1007/JHEP01(2011)094}{{\em JHEP} {\bf 1101} (2011)
   094},
\href{http://arxiv.org/abs/0809.2596}{{\tt arXiv:0809.2596 [hep-th]}}.

\bibitem{Erdmenger:2008rm}
J.~Erdmenger, M.~Haack, M.~Kaminski, and A.~Yarom, ``{Fluid dynamics of
  R-charged black holes},''
  \href{http://dx.doi.org/10.1088/1126-6708/2009/01/055}{{\em JHEP} {\bf 0901}
  (2009)  055},
\href{http://arxiv.org/abs/0809.2488}{{\tt arXiv:0809.2488 [hep-th]}}.

\bibitem{Bhattacharyya:2008ji}
S.~Bhattacharyya, R.~Loganayagam, S.~Minwalla, S.~Nampuri, S.~P. Trivedi, {\em
  et al.}, ``{Forced Fluid Dynamics from Gravity},''
  \href{http://dx.doi.org/10.1088/1126-6708/2009/02/018}{{\em JHEP} {\bf 0902}
  (2009)  018},
\href{http://arxiv.org/abs/0806.0006}{{\tt arXiv:0806.0006 [hep-th]}}.

\bibitem{Camps:2010br}
J.~Camps, R.~Emparan, and N.~Haddad, ``{Black Brane Viscosity and the
  Gregory-Laflamme Instability},''
  \href{http://dx.doi.org/10.1007/JHEP05(2010)042}{{\em JHEP} {\bf 1005} (2010)
   042},
\href{http://arxiv.org/abs/1003.3636}{{\tt arXiv:1003.3636 [hep-th]}}.

\bibitem{Gath:2013qya}
J.~Gath and A.~V. Pedersen, ``{Viscous Asymptotically Flat Reissner-Nordstr\"om
  Black Branes},''
\href{http://arxiv.org/abs/1302.5480}{{\tt arXiv:1302.5480 [hep-th]}}.

\bibitem{Harmark:2007md}
T.~Harmark, V.~Niarchos, and N.~A. Obers, ``{Instabilities of black strings and
  branes},'' \href{http://dx.doi.org/10.1088/0264-9381/24/8/R01}{{\em
  Class.Quant.Grav.} {\bf 24} (2007)  R1--R90},
\href{http://arxiv.org/abs/hep-th/0701022}{{\tt arXiv:hep-th/0701022
  [hep-th]}}.

\bibitem{Gregory:1993vy}
R.~Gregory and R.~Laflamme, ``{Black strings and p-branes are unstable},''
  \href{http://dx.doi.org/10.1103/PhysRevLett.70.2837}{{\em Phys.Rev.Lett.}
  {\bf 70} (1993)  2837--2840},
\href{http://arxiv.org/abs/hep-th/9301052}{{\tt arXiv:hep-th/9301052
  [hep-th]}}.

\bibitem{Gregory:1994bj}
R.~Gregory and R.~Laflamme, ``{The Instability of charged black strings and
  p-branes},'' \href{http://dx.doi.org/10.1016/0550-3213(94)90206-2}{{\em
  Nucl.Phys.} {\bf B428} (1994)  399--434},
\href{http://arxiv.org/abs/hep-th/9404071}{{\tt arXiv:hep-th/9404071
  [hep-th]}}.

\bibitem{Kol:2002xz}
B.~Kol, ``{Topology change in general relativity, and the black hole black
  string transition},''
  \href{http://dx.doi.org/10.1088/1126-6708/2005/10/049}{{\em JHEP} {\bf 0510}
  (2005)  049},
\href{http://arxiv.org/abs/hep-th/0206220}{{\tt arXiv:hep-th/0206220
  [hep-th]}}.

\bibitem{Emparan:2009at}
R.~Emparan, T.~Harmark, V.~Niarchos, and N.~A. Obers, ``{Essentials of
  Blackfold Dynamics},'' \href{http://dx.doi.org/10.1007/JHEP03(2010)063}{{\em
  JHEP} {\bf 1003} (2010)  063},
\href{http://arxiv.org/abs/0910.1601}{{\tt arXiv:0910.1601 [hep-th]}}.

\bibitem{Emparan:2011hg}
R.~Emparan, T.~Harmark, V.~Niarchos, and N.~A. Obers, ``{Blackfolds in
  Supergravity and String Theory},''
  \href{http://dx.doi.org/10.1007/JHEP08(2011)154}{{\em JHEP} {\bf 1108} (2011)
   154},
\href{http://arxiv.org/abs/1106.4428}{{\tt arXiv:1106.4428 [hep-th]}}.

\bibitem{Reall:2001ag}
H.~S. Reall, ``{Classical and thermodynamic stability of black branes},''
  \href{http://dx.doi.org/10.1103/PhysRevD.64.044005}{{\em Phys.Rev.} {\bf D64}
  (2001)  044005},
\href{http://arxiv.org/abs/hep-th/0104071}{{\tt arXiv:hep-th/0104071
  [hep-th]}}.

\bibitem{Bostock:2004mg}
P.~Bostock and S.~F. Ross, ``{Smeared branes and the Gubser-Mitra
  conjecture},'' \href{http://dx.doi.org/10.1103/PhysRevD.70.064014}{{\em
  Phys.Rev.} {\bf D70} (2004)  064014},
\href{http://arxiv.org/abs/hep-th/0405026}{{\tt arXiv:hep-th/0405026
  [hep-th]}}.

\bibitem{Aharony:2004ig}
O.~Aharony, J.~Marsano, S.~Minwalla, and T.~Wiseman, ``{Black hole-black string
  phase transitions in thermal 1+1 dimensional supersymmetric Yang-Mills theory
  on a circle},'' \href{http://dx.doi.org/10.1088/0264-9381/21/22/010}{{\em
  Class.Quant.Grav.} {\bf 21} (2004)  5169--5192},
\href{http://arxiv.org/abs/hep-th/0406210}{{\tt arXiv:hep-th/0406210
  [hep-th]}}.

\bibitem{Kovtun:2003wp}
P.~Kovtun, D.~T. Son, and A.~O. Starinets, ``{Holography and hydrodynamics:
  Diffusion on stretched horizons},''
  \href{http://dx.doi.org/10.1088/1126-6708/2003/10/064}{{\em JHEP} {\bf 0310}
  (2003)  064},
\href{http://arxiv.org/abs/hep-th/0309213}{{\tt arXiv:hep-th/0309213
  [hep-th]}}.

\bibitem{Kovtun:2004de}
P.~Kovtun, D.~Son, and A.~Starinets, ``{Viscosity in strongly interacting
  quantum field theories from black hole physics},''
  \href{http://dx.doi.org/10.1103/PhysRevLett.94.111601}{{\em Phys.Rev.Lett.}
  {\bf 94} (2005)  111601},
\href{http://arxiv.org/abs/hep-th/0405231}{{\tt arXiv:hep-th/0405231
  [hep-th]}}.

\bibitem{Buchel:2003tz}
A.~Buchel and J.~T. Liu, ``{Universality of the shear viscosity in
  supergravity},'' \href{http://dx.doi.org/10.1103/PhysRevLett.93.090602}{{\em
  Phys.Rev.Lett.} {\bf 93} (2004)  090602},
\href{http://arxiv.org/abs/hep-th/0311175}{{\tt arXiv:hep-th/0311175
  [hep-th]}}.

\bibitem{Buchel:2007mf}
A.~Buchel, ``{Bulk viscosity of gauge theory plasma at strong coupling},''
  \href{http://dx.doi.org/10.1016/j.physletb.2008.03.069}{{\em Phys.Lett.} {\bf
  B663} (2008)  286--289},
\href{http://arxiv.org/abs/0708.3459}{{\tt arXiv:0708.3459 [hep-th]}}.

\bibitem{Emparan:2013ila}
R.~Emparan, V.~E. Hubeny, and M.~Rangamani, ``{Effective hydrodynamics of black
  D3-branes},'' \href{http://dx.doi.org/10.1007/JHEP06(2013)035}{{\em JHEP}
  {\bf 1306} (2013)  035},
\href{http://arxiv.org/abs/1303.3563}{{\tt arXiv:1303.3563 [hep-th]}}.

\bibitem{Bredberg:2010ky}
I.~Bredberg, C.~Keeler, V.~Lysov, and A.~Strominger, ``{Wilsonian Approach to
  Fluid/Gravity Duality},''
  \href{http://dx.doi.org/10.1007/JHEP03(2011)141}{{\em JHEP} {\bf 1103} (2011)
   141},
\href{http://arxiv.org/abs/1006.1902}{{\tt arXiv:1006.1902 [hep-th]}}.

\bibitem{Bredberg:2011jq}
I.~Bredberg, C.~Keeler, V.~Lysov, and A.~Strominger, ``{From Navier-Stokes To
  Einstein},'' \href{http://dx.doi.org/10.1007/JHEP07(2012)146}{{\em JHEP} {\bf
  1207} (2012)  146},
\href{http://arxiv.org/abs/1101.2451}{{\tt arXiv:1101.2451 [hep-th]}}.

\bibitem{Caldarelli:2012hy}
M.~M. Caldarelli, J.~Camps, B.~Gouteraux, and K.~Skenderis, ``{AdS/Ricci-flat
  correspondence and the Gregory-Laflamme instability},''
\href{http://arxiv.org/abs/1211.2815}{{\tt arXiv:1211.2815 [hep-th]}}.

\bibitem{Caldarelli:2013aaa}
M.~M. Caldarelli, J.~Camps, B.~Goutéraux, and K.~Skenderis, ``{AdS/Ricci-flat
  correspondence},'' \href{http://dx.doi.org/10.1007/JHEP04(2014)071}{{\em
  JHEP} {\bf 1404} (2014)  071},
\href{http://arxiv.org/abs/1312.7874}{{\tt arXiv:1312.7874 [hep-th]}}.

\bibitem{Bhattacharyya:2008mz}
S.~Bhattacharyya, R.~Loganayagam, I.~Mandal, S.~Minwalla, and A.~Sharma,
  ``{Conformal Nonlinear Fluid Dynamics from Gravity in Arbitrary
  Dimensions},'' \href{http://dx.doi.org/10.1088/1126-6708/2008/12/116}{{\em
  JHEP} {\bf 0812} (2008)  116},
\href{http://arxiv.org/abs/0809.4272}{{\tt arXiv:0809.4272 [hep-th]}}.

\bibitem{Haack:2008cp}
M.~Haack and A.~Yarom, ``{Nonlinear viscous hydrodynamics in various dimensions
  using AdS/CFT},'' \href{http://dx.doi.org/10.1088/1126-6708/2008/10/063}{{\em
  JHEP} {\bf 0810} (2008)  063},
\href{http://arxiv.org/abs/0806.4602}{{\tt arXiv:0806.4602 [hep-th]}}.

\bibitem{Argurio:1997gt}
R.~Argurio, F.~Englert, and L.~Houart, ``{Intersection rules for p-branes},''
  \href{http://dx.doi.org/10.1016/S0370-2693(97)00205-0}{{\em Phys.Lett.} {\bf
  B398} (1997)  61--68},
\href{http://arxiv.org/abs/hep-th/9701042}{{\tt arXiv:hep-th/9701042
  [hep-th]}}.

\bibitem{Argurio:1997nh}
R.~Argurio, ``{Intersection rules and open branes},''
\href{http://arxiv.org/abs/hep-th/9712170}{{\tt arXiv:hep-th/9712170
  [hep-th]}}.

\bibitem{Peet:2000hn}
A.~W. Peet, ``{TASI lectures on black holes in string theory},''
\href{http://arxiv.org/abs/hep-th/0008241}{{\tt arXiv:hep-th/0008241
  [hep-th]}}.

\bibitem{Huq:1983im}
M.~Huq and M.~Namazie, ``{{Kaluza-Klein} Supergravity in Ten-dimensions},''
\href{http://dx.doi.org/10.1088/0264-9381/2/3/007}{{\em Class.Quant.Grav.} {\bf
  2} (1985)  293}.

\bibitem{Giani:1984wc}
F.~Giani and M.~Pernici, ``{N=2 supergravity in ten-dimensions},''
\href{http://dx.doi.org/10.1103/PhysRevD.30.325}{{\em Phys.Rev.} {\bf D30}
  (1984)  325--333}.

\bibitem{Campbell1984112}
I.~Campbell and P.~West, ``N = 2, d = 10 non-chiral supergravity and its
  spontaneous compactification,''
  \href{http://dx.doi.org/http://dx.doi.org/10.1016/0550-3213(84)90388-2}{{\em
  Nuclear Physics B} {\bf 243} (1984) no.~1, 112 -- 124}.

\bibitem{Cremmer:1978km}
E.~Cremmer, B.~Julia, and J.~Scherk, ``{Supergravity Theory in
  Eleven-Dimensions},''
\href{http://dx.doi.org/10.1016/0370-2693(78)90894-8}{{\em Phys.Lett.} {\bf
  B76} (1978)  409--412}.

\bibitem{Tseytlin:1996cc}
A.~A. Tseytlin, ``{Composite black holes in string theory},''
\href{http://arxiv.org/abs/gr-qc/9608044}{{\tt arXiv:gr-qc/9608044 [gr-qc]}}.

\bibitem{Caldarelli:2010xz}
M.~M. Caldarelli, R.~Emparan, and B.~Van~Pol, ``{Higher-dimensional Rotating
  Charged Black Holes},'' \href{http://dx.doi.org/10.1007/JHEP04(2011)013}{{\em
  JHEP} {\bf 1104} (2011)  013},
\href{http://arxiv.org/abs/1012.4517}{{\tt arXiv:1012.4517 [hep-th]}}.

\bibitem{Armas:2011uf}
J.~Armas, J.~Camps, T.~Harmark, and N.~A. Obers, ``{The Young Modulus of Black
  Strings and the Fine Structure of Blackfolds},''
  \href{http://dx.doi.org/10.1007/JHEP02(2012)110}{{\em JHEP} {\bf 1202} (2012)
   110},
\href{http://arxiv.org/abs/1110.4835}{{\tt arXiv:1110.4835 [hep-th]}}.

\bibitem{Emparan:2011br}
R.~Emparan, ``{Blackfolds},''
\href{http://arxiv.org/abs/1106.2021}{{\tt arXiv:1106.2021 [hep-th]}}.

\bibitem{Gubser:2000ec}
S.~S. Gubser and I.~Mitra, ``{Instability of charged black holes in Anti-de
  Sitter space},''
\href{http://arxiv.org/abs/hep-th/0009126}{{\tt arXiv:hep-th/0009126
  [hep-th]}}.

\bibitem{Gubser:2000mm}
S.~S. Gubser and I.~Mitra, ``{The Evolution of unstable black holes in anti-de
  Sitter space},'' \href{http://dx.doi.org/10.1088/1126-6708/2001/08/018}{{\em
  JHEP} {\bf 0108} (2001)  018},
\href{http://arxiv.org/abs/hep-th/0011127}{{\tt arXiv:hep-th/0011127
  [hep-th]}}.

\bibitem{Ross:2005vh}
S.~F. Ross and T.~Wiseman, ``{Smeared D0 charge and the Gubser-Mitra
  conjecture},'' \href{http://dx.doi.org/10.1088/0264-9381/22/14/006}{{\em
  Class.Quant.Grav.} {\bf 22} (2005)  2933--2946},
\href{http://arxiv.org/abs/hep-th/0503152}{{\tt arXiv:hep-th/0503152
  [hep-th]}}.

\bibitem{Harmark:2005jk}
T.~Harmark, V.~Niarchos, and N.~A. Obers, ``{Instabilities of near-extremal
  smeared branes and the correlated stability conjecture},''
  \href{http://dx.doi.org/10.1088/1126-6708/2005/10/045}{{\em JHEP} {\bf 0510}
  (2005)  045},
\href{http://arxiv.org/abs/hep-th/0509011}{{\tt arXiv:hep-th/0509011
  [hep-th]}}.

\bibitem{Gouteraux:2011qh}
B.~Gouteraux, J.~Smolic, M.~Smolic, K.~Skenderis, and M.~Taylor, ``{Holography
  for Einstein-Maxwell-dilaton theories from generalized dimensional
  reduction},'' \href{http://dx.doi.org/10.1007/JHEP01(2012)089}{{\em JHEP}
  {\bf 1201} (2012)  089},
\href{http://arxiv.org/abs/1110.2320}{{\tt arXiv:1110.2320 [hep-th]}}.

\bibitem{Kanitscheider:2009as}
I.~Kanitscheider and K.~Skenderis, ``{Universal hydrodynamics of non-conformal
  branes},'' \href{http://dx.doi.org/10.1088/1126-6708/2009/04/062}{{\em JHEP}
  {\bf 0904} (2009)  062},
\href{http://arxiv.org/abs/0901.1487}{{\tt arXiv:0901.1487 [hep-th]}}.

\bibitem{Gregory:2000gf}
R.~Gregory, ``{Black string instabilities in Anti-de Sitter space},''
  \href{http://dx.doi.org/10.1088/0264-9381/17/18/103}{{\em Class.Quant.Grav.}
  {\bf 17} (2000)  L125--L132},
\href{http://arxiv.org/abs/hep-th/0004101}{{\tt arXiv:hep-th/0004101
  [hep-th]}}.

\bibitem{Buchel:2011uj}
A.~Buchel, ``{Violation of the holographic bulk viscosity bound},''
  \href{http://dx.doi.org/10.1103/PhysRevD.85.066004}{{\em Phys.Rev.} {\bf D85}
  (2012)  066004},
\href{http://arxiv.org/abs/1110.0063}{{\tt arXiv:1110.0063 [hep-th]}}.

\bibitem{DiDato:2013cla}
A.~Di~Dato, ``{Kaluza-Klein reduction of relativistic fluids and their gravity
  duals},'' \href{http://dx.doi.org/10.1007/JHEP12(2013)087}{{\em JHEP} {\bf
  1312} (2013)  087},
\href{http://arxiv.org/abs/1307.8365}{{\tt arXiv:1307.8365 [hep-th]}}.

\bibitem{Hirayama:2002hn}
T.~Hirayama, G.~Kang, and Y.~Lee, ``{Classical stability of charged black
  branes and the Gubser-Mitra conjecture},''
  \href{http://dx.doi.org/10.1103/PhysRevD.67.024007}{{\em Phys.Rev.} {\bf D67}
  (2003)  024007},
\href{http://arxiv.org/abs/hep-th/0209181}{{\tt arXiv:hep-th/0209181
  [hep-th]}}.

\bibitem{Erdmenger:2014jba}
J.~Erdmenger, M.~Rangamani, S.~Steinfurt, and H.~Zeller, ``{Hydrodynamic
  Regimes of Spinning Black D3-Branes},''
\href{http://arxiv.org/abs/1412.0020}{{\tt arXiv:1412.0020 [hep-th]}}.

\bibitem{genq}
J.~Gath and A.~V. Pedersen, ``{Work in progress},'' {\em Forthcoming} (2015)  .

\end{thebibliography}\endgroup

\end{document}